\documentstyle[preprint,aps,psfig]{revtex}
\tighten

\begin{document}
\preprint{Preprint Number:
\parbox[t]{50mm}{ADP-97-10/T248 \\
		 hep-ph/9705298}  }
\draft
%_______________________ Title, Authors ____________________________________
\title{Solving the Bethe-Salpeter equation for bound states of scalar \\
       theories in Minkowski space}
\author{
  Kensuke Kusaka$^1$\footnote{JSPS Research Fellow}
  Ken Simpson$^2$
and
  Anthony G. Williams$^{2,3}$
  \vspace*{2mm} }

\address{
    ${}^1$ Department of Physics, Tokyo Metropolitan University, \\
    Minami-Osawa 1-1, Hachioji-shi, Tokyo 192-03, Japan\\
    \vspace*{2mm}
    ${}^2$ Department of Physics and Mathematical Physics, \\
    University of Adelaide, South Australia 5005, Australia\\
    \vspace*{2mm}
    ${}^3$ Special Research Centre for the Subatomic Structure of Matter,\\
    University of Adelaide, South Australia 5005, Australia\\
    \vspace*{4mm}
    {\rm e-mail:} kkusaka@phys.metro-u.ac.jp,
    ksimpson@physics.adelaide.edu.au, awilliam@physics.adelaide.edu.au
}
%\date{}
%
\maketitle

\begin{abstract}
We apply the perturbation theory integral representation (PTIR)
to solve for the bound state Bethe-Salpeter (BS) vertex
for an {\it arbitrary} scattering kernel, without the need for
any Wick rotation.  The results derived are applicable to any scalar
field theory (without derivative coupling).
It is shown that solving directly for the BS vertex, rather than the
BS amplitude, has several major advantages, notably its relative simplicity
and superior numerical accuracy.  In order to illustrate the generality
of the approach we obtain numerical solutions using this formalism
for a number of scattering kernels, including cases where the Wick rotation
is not possible. 
\end{abstract}


\section{Introduction}

We present here an
improved approach to the solution of the scalar-scalar Bethe-Salpeter (BS)
equation directly in Minkowski space, utilising the Perturbation
Theory Integral Representation (PTIR) of Nakanishi~\cite{Nakanishi_graph}.
The PTIR is a generalised spectral representation for $n$-point Green's
functions in Quantum Field Theory.

This work extends and improves
earlier work which applied the PTIR approach to the BS amplitude
\cite{BS_amplitude}.  Here we formulate a real integral equation for the BS
vertex. This considerably simplifies the expression for the kernel function
relative to those obtained for the BS amplitude \cite{BS_amplitude}.
In particular, some singularity structures which were present in the
kernel of the BS amplitude equation due to the
constituent particle propagators are absent in the corresponding
expression for the BS vertex.  Consequently, it is much simpler to implement
the problem numerically, and we therefore do not encounter previous
difficulties with residual numerical noise. We have checked that our
numerical results are in agreement with those obtained in Euclidean space
by other authors (for example, Linden and Mitter~\cite{L+M}, or more recently,
Nieuwenhuis and Tjon~\cite{Tjon1,Tjon2}).  In particular, with sufficient
computer time we have seen no limit to the accuracy that can be achieved
with our formalism.  We can routinely obtain four figure accuracy on a
workstation.

In this work we will deal exclusively with scalar theories. 
For simplicity we will consider here bound states with equal-mass
constituents, although it is easy, and desirable in many applications, 
to generalise the approach to unequal-mass constituents.

\begin{figure}[hbt]  
\centering{\
     \psfig{angle=270,figure=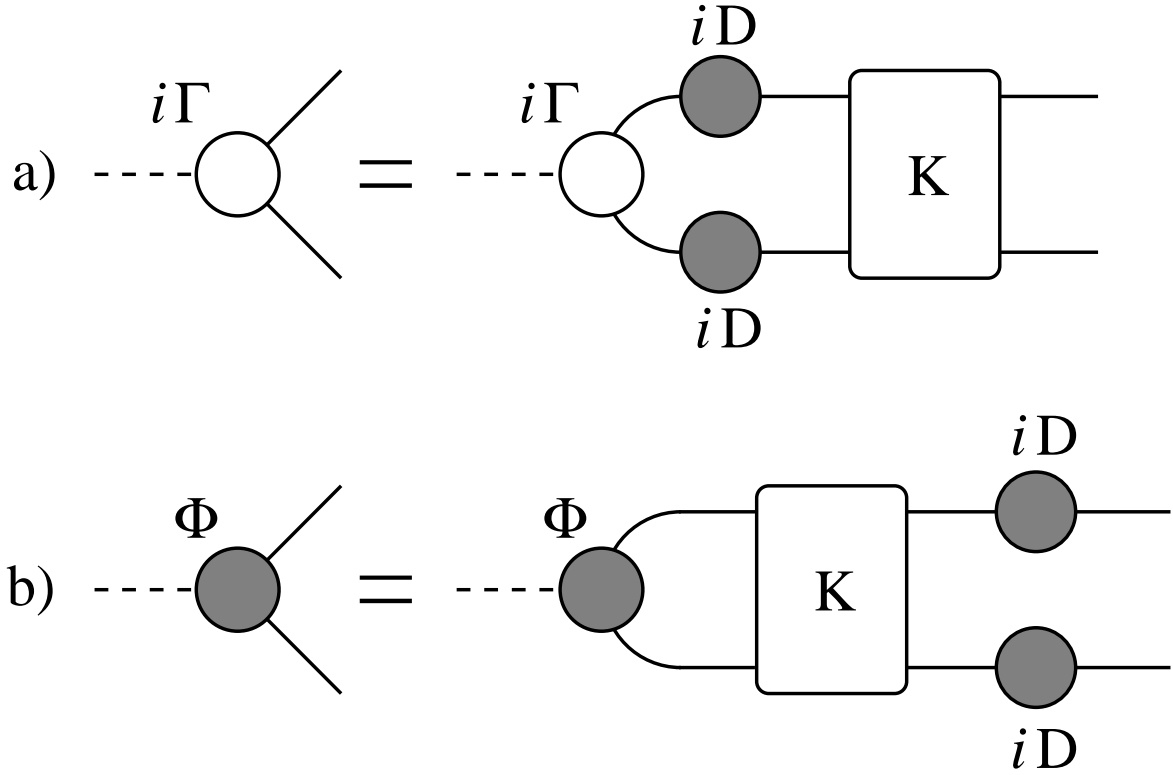,height=12.5cm} }
\protect\parbox{130mm}{\caption{Diagrammatic representation of the
Bethe-Salpeter equation for (a) the BS vertex ($\Gamma$) and 
(b) the BS amplitude ($\Phi$). The fully dressed constituent particle
propagator is denoted by $D$ and $K$ is the scattering kernel for
the constituents.}
\label{fig_BSE}
}
\end{figure}

We illustrate the BS equation for a scalar
theory in Fig.~\ref{fig_BSE}, where $\Phi(p,P)$
is the BS amplitude, $P\equiv p_1 + p_2$ is the total four-momentum of
the bound state
and $p\equiv \eta_2 p_1 -\eta_1 p_2$ is the relative four-momentum for
the two scalar constituents.  We have then
$M=\sqrt{P^2}$ for the bound state mass and also
$\eta_1+\eta_2=1$, but otherwise
the choice of the two positive real numbers $\eta_1$ and
$\eta_2$ is arbitrary. As in Ref.~\cite{BS_amplitude} we choose here
$\eta_1=\eta_2$=1/2.

The renormalised constituent scalar propagators are
$D(p_{1,2}^2)$ and
$K(p,q;P)$ is the renormalised scattering kernel.  For example, in simple
ladder approximation in a $\phi^2\sigma$ model we would have
$K(p,q;P)=(ig)(iD_\sigma([p-q]^2))(ig)$, where
$D_\sigma(p^2)=1/(p^2-m_\sigma^2+i\epsilon)$ and $m_\sigma$ is the
$\sigma$-particle mass.  Note that the corresponding
proper (i.e, one-particle irreducible) vertex for the bound state
is related to the BS amplitude by $\Phi=(iD)(i\Gamma)(iD)$. 

We follow standard conventions in our definitions of quantities,
(see Ref.~\cite{B+D,I+Z} and also, e.g., \cite{TheReview}).  Thus the
BS equation
for any scalar theory can be written as
\begin{equation}
i\Gamma(p_1,p_2) = \int{d^4q\over(2\pi)^4}
        (iD(q_1^2))~
        (i\Gamma(q_1,q_2))~(iD(q_2^2))~K(p,q;P)\, ,
	\label{BSE_Gamma}
\end{equation}
where similarly to $p_1$ and $p_2$ we have defined
$q_1=\eta_1 P+q$ and $q_2=\eta_2 P-q$.
Equivalently in terms of the BS amplitude we can write
\begin{eqnarray}
  D(p_1^2)^{-1}~\Phi(p,P)~D(p_2^2)^{-1}
        &=&-\int{d^4q\over(2\pi)^4}~\Phi(q,P)~K(p,q;P)
	\label{BSE_K} \\
        &\equiv&~\int{d^4q\over(2\pi)^4 i}~\Phi(q,P)~I(p,q;P)\, ,
	\label{BSE}
\end{eqnarray}
where the kernel function defined by $I(p,q;P)\equiv-iK(p,q;P)$ is the form
typically used by Nakanishi \cite{Nakanishi_survey}.  In ladder approximation
for a $\phi^2\sigma$ model we see for example that
$I(p,q;P)=g^2/(m_\sigma^2-(p-q)^2-i\epsilon)$. In this treatment we will
solve the vertex version of the BS equation, {\it i.e.} Eq. (\ref{BSE_Gamma}),
for an {\it arbitrary} scattering kernel. 

\section{PTIR for Scalar Theories}
\label{Representation}

In contrast to the approach in Ref.~\cite{BS_amplitude}, we will begin with the 
Bethe-Salpeter equation for the bound-state vertex, Eq. (\ref{BSE_Gamma}),
rather than the equivalent equation for the amplitude, Eq. (\ref{BSE}),
and derive a real integral equation for the BS vertex.
To do this we require spectral representations for the vertex, for
the $\phi$ propagator, and for the scattering kernel $I$ of 
Eq. (\ref{BSE_Gamma}). The renormalised $\phi$ propagator may be written as
\begin{equation}
        D(q)=-\left({1 \over m^2 - q^2 -i\epsilon}
                + \int_{(m+\mu)^2}^\infty d\alpha
                \frac{\rho_\phi(\alpha)}{\alpha-q^2-i\epsilon}\right),
        \label{phi_prop}
\end{equation}
where $\rho_\phi(\alpha)$ is the renormalised spectral function.
Note that $\rho_\phi(\alpha)\ge 0$, (see, e.g.,
Ref~\cite{B+D}).

The Bethe-Salpeter amplitude $\Phi(p,P)$ for the bound state
of two $\phi$-particles having the total momentum $P\equiv p_1 +p_2$ and
relative momentum $p\equiv (\eta_2p_1-\eta_1p_2)$ can be defined as
\begin{equation}
	\langle 0|T\phi(x_1)\phi(x_2)|P\rangle
	= e^{-iP\cdot X}\langle 0|T\phi(\eta_2 x)\phi(-\eta_1 x)|P\rangle
	= e^{-iP\cdot X}
	\int{d^4p\over (2\pi)^4} e^{-ip\cdot x} \Phi(p,P)\;,
\label{BS_amp}
\end{equation}
where the fields for the scalar constituents are denoted by $\phi$, and
where we have made use of the translational invariance of the BS amplitude.
Following the conventions of Itzykson and Zuber~\cite{I+Z} (e.g.,
pp.~481-487), we define centre-of-momentum and relative coordinates
$X\equiv \eta_1x_1+\eta_2x_2$ and $x\equiv x_1-x_2$ such that
$x_1=X+\eta_2x$, $x_2=X-\eta_1x$, and
$P\cdot X + p\cdot x = p_1\cdot x_1 + p_2\cdot x_2$.  

Equivalently to Eq.(\ref{BS_amp}), we can write
\begin{equation}
	\Phi(p,P)
	=e^{iP\cdot X}\int d^4x e^{ip\cdot x}
		\langle 0|T\phi(x_1)\phi(x_2)|P\rangle
	=\int d^4x e^{ip\cdot x}
		\langle 0|T\phi(\eta_2 x)\phi(-\eta_1 x)|P\rangle\;.
\label{BS_FT}
\end{equation}
Note that the bound states are normalised such that
$\langle P|P'\rangle=2\omega_P(2\pi)^3\delta^3(\vec P'-\vec P)$,
where $\omega_P\equiv ({\vec P}^2+M^2)^{1/2}$ with $M$ the bound state mass.
For a positive energy bound state we must have $P^2=M^2$,
$0 < P^2 \le (2m)^2$, and $P^0 > 0$.
The normalization condition for the BS amplitude is given by
\begin{equation}
\int \frac{d^4p}{(2\pi)^4}\int \frac{d^4q}{(2\pi)^4}
	\bar\Phi(q,P)\frac{\partial}{\partial P_\mu}
	\left\{D^{-1}(p_1^2)D^{-1}(p_2^2)(2\pi)^4\delta^4(p-q)
	+K(p,q;P)\right\}\Phi(p,P)=2iP^\mu\;,
\label{normalization}
\end{equation}
where the conjugate BS amplitude $\bar\Phi(p,P)$ is defined by
\begin{equation}
	\bar\Phi(p,P)=e^{-iP\cdot X}\int d^4x e^{-ip\cdot x}
		\langle P|T\phi^\dagger(x_1)\phi^\dagger(x_2)|0\rangle
	=\int d^4x e^{-ip\cdot x}
	\langle P|T\phi^\dagger(\eta_2 x)\phi^\dagger(-\eta_1 x)|0\rangle\;.
\end{equation}

\subsection{PTIR for Scattering Kernel}
\label{section_PTIR}
The scattering kernel $I(p,q;P)\equiv -iK(p,q;P)$ describes the process
$\phi\phi \rightarrow \phi\phi$, where
$p$ and $q$ are the initial and final relative momenta
respectively.  It is given by the infinite series of Feynman diagrams
which are two-particle irreducible
with respect to the initial and final pairs of constituent $\phi$
particles.
For purely scalar theories without derivative coupling we have
the formal expression for the full
renormalised scattering kernel~\cite{Nakanishi_graph}
\begin{eqnarray}
    I(p,q;P) &=&
    \int\limits_{0}^{\infty } d\gamma\int\limits_{\Omega } d\vec\xi
     \left\{ \frac{ \rho_{st} (\gamma,\vec\xi) }
    { \gamma - \left[ \sum_{i=1}^{4}\xi_iq_i^2+\xi_5s+\xi_6t
     \right]-i\epsilon }
    \right.
    + \frac{ \rho_{tu}(\gamma,\vec\xi) }
    { \gamma - \left[ \sum_{i=1}^{4}\xi_iq_i^2+\xi_5t+\xi_6u \right]
     -i\epsilon }
    \nonumber\\
    &  & \left. + \frac{\rho_{us}(\gamma,\vec\xi) }
    {\gamma-\left[\sum_{i=1}^{4}\xi_iq_i^2+\xi_5u+\xi_6s\right]-i\epsilon}
    \right\}\;,
    \label{krnl_PTIR}
\end{eqnarray}
where $q_i^2$ is the 4-momentum squared carried by $\phi_i$
and $s$,$t$ and $u$ are the usual Mandelstam variables.
The symbol $\Omega$ denotes the integral region of
$\xi_i$ such that $\Omega\equiv\{\xi_i \,| \,0\leq\xi_i\leq 1, \,
\sum\xi_i=1 \, (i=1,\dots, 6)\}$. The scattering kernel PTIR
can be re-written in a more compact
form as
\begin{eqnarray}
    I(p,q;P) &=&
    \sum_{\rm ch}\int\limits_{0}^{\infty } d\gamma\int\limits_{\Omega } d\vec\xi
     \frac{ \rho_{\rm ch} (\gamma,\vec\xi) }
    { \gamma - (a_{\rm ch}q^2 + b_{\rm ch}p.q +
      c_{\rm ch}p^2 + d_{\rm ch}P^2 + e_{\rm ch}q.P+
	f_{\rm ch}p.P ) -i\epsilon } ,
    \label{krnl_PTIR_sum}
\end{eqnarray}
where the subscript ``ch'' indicates which channel we are dealing with
(either $st$, $tu$ or $us$), and \{$a_{\rm ch},\ldots,f_{\rm ch}$\}
are linear combinations of the $\xi_i$ (see Appendix~\ref{apdx_a}).

For more general theories involving, e.g., fermions and/or derivative
couplings, the numerator of Eq.~(\ref{krnl_PTIR})
will also contain momenta in general.  Work is in progress
to extend the formalism to include cases, such as derivative coupling
and fermions, where momentum dependence exists in
the numerator.

To illustrate our approach we will present here results for three
choices of kernel: \\
	{\bf (a)} Scalar-scalar ladder model with massive scalar exchange:
	The simple $t$-channel one-$\sigma$-exchange kernel is given by
	\begin{equation}
		I(p,q;P)=\frac{g^2}{m_\sigma^2-(p-q)^2-i\epsilon}
		\label{pure_ladder_kernel}
	\end{equation}
	The BS equation with this kernel together with the perturbative
	constituent particle propagator $D^0$ is often referred to as the
	``scalar-scalar ladder model''\cite{Nakanishi_survey}. 
	Note that the kernel weight function is
	proportional to $g^2$ for this simple case. \\
	{\bf (b)} Dressed ladder model:
	In this instance we dress the propagator of the exchanged 
	$\sigma$ of case (a). The kernel then consists of the
	pole term as above, plus a piece which involves an integration
	(starting at a threshold of $4m^2$) over a mass parameter $\gamma$. \\
	{\bf (c)} Generalised kernel:
	A sum of the one-$\sigma$-exchange kernel
	Eq.~(\ref{pure_ladder_kernel})
	and a generalised kernel with fixed kernel parameter sets
	$\{\gamma^{(i)}, \vec\xi^{(i)}\}$.
	After the Wick rotation this kernel
	becomes complex due to the $p\cdot P$ and $q\cdot P$ terms, so that
	solving the BS amplitude as a function of Euclidean relative momentum
	would be very difficult in this case.

The Wick rotated BS equation for cases (a) and (b) has been studied
numerically\cite{L+M}.  We use the previous numerical results for these 
kernels as a check of our new technique and numerical
calculations.

\subsection{PTIR for BS vertex}
\label{sec_PTIR_BS}

As in Ref.~\cite{BS_amplitude}, we will use the following form of the
$s$-wave ($\ell=0$) BS vertex:
	\begin{equation}
	\Gamma(p,P)=\int\limits_{0}^{\infty}d\alpha
		\int\limits_{-1}^{1}dz
		\frac{\rho_n(\alpha,z)}
		{\left[m^2+\alpha-\left(p^2 + z p\cdot P+\frac{P^2}{4}\right)
			-i\epsilon\right]^{n}}\;,
	\label{vtexPTIR2}
	\end{equation}
with the following boundary condition for the weight function:
	\begin{equation}
	\lim_{\alpha\rightarrow\infty}
			\frac{\rho_n(\alpha,z)}{\alpha^{n-1}}=0\;,
	\label{bc_PTIR}
	\end{equation}
required to render the $\alpha$--integral finite. A partial integration
of Eq.(\ref{vtexPTIR2}) with respect to $\alpha$ will serve to demonstrate
that the positive integer $n$ is a dummy parameter, 
since it follows from such a step that
weight functions associated with successive values of $n$ are connected
by the following relation:
	\begin{equation}
		\rho_{n+1}(\alpha,z)=n\int\limits_{0}^\alpha
					d\alpha'\rho_{n}(\alpha',z)\;.
		\label{rln_wghtfnc}
	\end{equation}
Note that the larger the dummy integer parameter $n$, the smoother
the corresponding weight function. This is a particularly useful
observation since we have a numerical solution of the BS equation
in mind.

Using the same arguments as outlined in~\cite{BS_amplitude}, it can be shown 
that the vertex PTIR for bound states with non-zero angular momentum
$\ell$ in
an arbitrary frame of reference is
\begin{equation}
	\Gamma^{[\ell,\ell_z]}(p,P)={\cal Y}_\ell^{\ell_z}(\vec{p'})
		\int\limits_{0}^{\infty}d\alpha
		\int\limits_{-1}^{1}dz
		\frac{\rho^{[\ell]}_n(\alpha,z)}
		{\left[m^2+\alpha-\left(p^2 + z p\cdot P + P^2/4\right)
			-i\epsilon\right]^{n}}\;,
	\label{ptlw_BS_PTIR2}
\end{equation}
where $P$ is an arbitrary timelike 4-vector with $P^2=M^2$
and $p'=\Lambda^{-1}(P)p$.  The Lorentz transformation
$\Lambda(P)$ connects $P$ and the bound-state rest frame 4-vector
$P'=(M,\vec 0)$, i.e, $P = \Lambda(P) P'$. The quantity 
${\cal Y}_\ell^{\ell_z}(\vec{p'})$ is the solid harmonic of order
$\ell$, and may be written in the form $|\vec{p'}|^\ell 
Y_\ell^{\ell_z}(\hat{p}')$, where $Y_\ell^{\ell_z}$ is the ordinary 
spherical harmonic of order $\ell$, 
$\hat{p}'\equiv \vec{p'}/|\vec{p'}|$, and where $\vec{p'}$ is the
3-vector relative momentum in the bound state rest frame. It is relatively
straightforward to appreciate why Eq.~(\ref{ptlw_BS_PTIR2}) must be the
correct form for a scalar bound state with angular momentum $\ell$.
It follows from the self-reproducing property of the solid harmonics
(see Eq.~(\ref{solid_harm_prop}) in Appendix~\ref{apdx_b}) and from the
fact that in the bound state rest frame $\vec{p'}$ is the only available
three-vector.

In the following sections
we will study the BS equation Eq.~(\ref{BSE_Gamma}) in an arbitrary frame
in terms of this integral representation.
Note that the dummy parameter $n$ can always be taken sufficiently large
such that
the loop-momentum integral of the BS equation, Eq.~(\ref{BSE_Gamma}),
converges for any $\ell$ for which a bound state exists.

Before proceeding, we should address the issue of so-called
ghost states.
A bound state whose BS amplitude $\Phi(p,P)$ and 
equivalent vertex $\Gamma(p,P)$ are anti--symmetric 
under the transformation $P\cdot p \rightarrow - P\cdot p$ 
with fixed $p^{2}$ and $P^{2}$ has a negative norm and is called a 
``ghost''\cite{Nakanishi_survey}.  This symmetry corresponds to 
the one: $\bar z \rightarrow -\bar z$ in PTIR form. We do not
consider such states herein, as they are unphysical.

\section{BS Equation for the Weight Function}
\label{Integral}

In this section we will reformulate the BS equation 
Eq.~(\ref{BSE_Gamma})
as an integral equation in terms of the weight functions.  This is the
central result of this paper. We will very briefly describe the
procedure, and state our main results; the details of the 
derivation may be found in Appendix~\ref{apdx_b}, as may definitions
of the kernel and associated functions.

We proceed by combining, using Feynman parametrisation, the integral 
representations for the scattering kernel and vertex with the bare propagators
for the constituent $\phi$ particles. The procedure can easily be generalised
to include dressed constituent propagators if desired, but we do not exercise
this option here, for the sake of simplicity. After using the PTIR
representations for the BS kernel and vertex ({\it i.e.}, 
Eqs. (\ref{krnl_PTIR_sum}) and (\ref{ptlw_BS_PTIR2}))
in the right-hand side of the BS equation (Eq. (\ref{BSE_Gamma})),
and after performing Feynman parametrisation for the right-hand side,
the BS equation can be written as

\begin{eqnarray}
\Gamma^{[\ell,\ell_z]}(p,P)&=&{\cal Y}_\ell^{\ell_z}
	\int\limits_{-\infty}^{\infty}d\bar{\alpha}
	\int\limits_{-1}^{1}d\bar{z} \frac{1}{[F(\bar{\alpha},
	\bar{z};p,P)-i\epsilon]^n}\nonumber \\
&\times&\int\limits_{0}^{\infty} d\alpha\int\limits_{-1}^{1}dz 
	\frac{\rho_n^{[\ell]}(\alpha,z)}{\alpha^n}\;
	\bar{\alpha}^n[\lambda\;
	{}^{\rm tot}\!{\cal K}_n^{[\ell]}(\bar{\alpha},\bar{z};\alpha,z)],
\label{new_vert}
\end{eqnarray}
where $F(\alpha,z;p,P)\equiv m^2+\alpha-(q^2+zp\cdot P + \frac{1}{4}P^2)$
is a convenient shorthand notation. We have defined an ``eigenvalue''
$\lambda\equiv g^2/(4\pi)^2$, which we will use in our numerical work (see
Appendix~\ref{apdx_0}). This has simply been factored out of the scattering
kernel for convenience and for ease of comparison with other calculations
in the ladder limit. The total kernel function 
${}^{\rm tot}\!{\cal K}_n^{[\ell]}$ is defined in Appendix~\ref{apdx_b} 
(see Eqs.~(\ref{tot_kernel}) and~(\ref{kernel})) and its structure
is discussed in Appendix~\ref{Singularity} with particualr attention
paid to any potential singularities.

Comparing Eq.~(\ref{new_vert}) with Eq.~(\ref{ptlw_BS_PTIR2}),
and using the uniqueness theorem of PTIR\cite{Nakanishi_graph},
we finally 
obtain the following integral equation for $\rho^{[\ell]}_n(\alpha,z)$:
\begin{equation}
	\frac{1}{\lambda}
	\frac{\rho^{[\ell]}_n(\bar\alpha,\bar z)}{\bar\alpha^n}=
	  \int\limits_{0}^{\infty}d\alpha
	\int\limits_{-1}^{1}dz \,
	{}^{\rm tot}\!{\cal K}^{[\ell]}_n(\bar\alpha,\bar z;\alpha,z)
	\frac{\rho^{[\ell]}_n(\alpha,z)}{\alpha^n}.
	\label{eqn_wght}
\end{equation}
This equation is the central result of this work.
Note that in Eq.(\ref{eqn_wght})
we are solving for $\rho_n^{[\ell]}/\alpha^n$; this is for reasons of
convenience for our numerical treatment of the BS equation. 
\vskip 12pt
\noindent
{\bf Summary:} Since the weight functions $\rho_{\rm ch}(\gamma,\vec\xi)$
for the scattering kernel are real functions by their construction,
the total kernel function 
${}^{\rm tot}\!{\cal K}^{[\ell]}_n(\bar\alpha,\bar z;\alpha,z)$
is real, so that
Eq.~(\ref{eqn_wght}) is a real integral equation in 
two variables $\alpha$ and $z$.  Thus we have transformed the
BS equation, which is a singular integral equation of
complex distributions, into a real integral equation which is
frame-independent.
Once one solves for the BS vertex weight function, the BS vertex
and the BS amplitude
can be written down in an arbitrary frame.
This is clearly advantageous for applications of the BS amplitude to
relativistic problems.

\section{Numerical Results}
\label{Results}

In this section we present numerical solutions for the BS
vertex for bound states in scalar theories using
Eq.(\ref{eqn_wght}) for
three simple choices of scattering kernel: 
\vspace{4mm} \\
(a) pure ladder kernel with massive scalar exchange, \\
(b) dressed ladder kernel with pole term as in (a), and \\ 
(c) a generalised kernel combined with the pure ladder
kernel of (a). 
\vspace{4mm}
 
The scattering kernel (a), i.e., the one-$\sigma$-exchange
kernel depending only on $t=(p-q)^2$, is given by
Eq.({\ref{pure_ladder_kernel}}).
This corresponds to choosing for the kernel in Eq.~(\ref{krnl_PTIR}) say:
$\rho_{tu}=\rho_{us}=0$ and in the $st$-channel $\gamma=m_\sigma^2$,
and $a_{st}=c_{st}=1$, $b_{st}=-2$, $d_{st}=e_{st}=f_{st}=0$,
[c.f., Eq.~(\ref{krnl_PTIR_sum})], which amounts to
choosing $\rho_{st}$ to be an appropriate product of
$\delta$-functions multiplied by $g^2$.
In the pure ladder case it is
convenient (and traditional) to factorise out
the coupling constant $g^2$ and a factor of $(4\pi)^2$, by defining
the ``eigenvalue'' $\lambda=g^2/(4\pi)^2$ \cite{Nakanishi63,Sato,L+M}.
Thus it is usual to fix the bound state mass $P^2$ and then to solve for
the coupling $g^2$, which is what we have done here. 

In general (and for example, for the dressed ladder kernel) the kernel
depends on higher powers of the coupling than $g^2$; in such instances 
we update the running value of $\lambda$ in an appropriate way during the 
iteration process (see Appendix \ref{apdx_0}).

Since the BSE is a homogeneous integral equation and we are only
interested at present in extracting the coupling at fixed bound state
mass, the choice of normalisation is unimportant provided that it is 
fixed in some reasonable way.

The numerical solution of the vertex BSE is performed by choosing a 
suitable grid of $\alpha$ and $z$ values, making an initial guess
for the vertex weight function, and then iterating the
integral equation~(\ref{eqn_wght}) to covergence. One subtle point is
that integrable square-root singularities may occur and must be
appropriately handled numerically (see Appendices~\ref{apdx_0} 
and~\ref{Singularity} for details). By optimising the choice of
grid and increasing the number of grid points it was straightforward
to increase the accuracy of the solutions to a relative error
of 1 part in 10$^4$ and beyond. Further details of the numerical 
procedure used are given in Appendix~\ref{apdx_0}.

\subsection{Pure Ladder Kernel}

\begin{figure}[hbt]
\centering{\
     \psfig{angle=0,figure=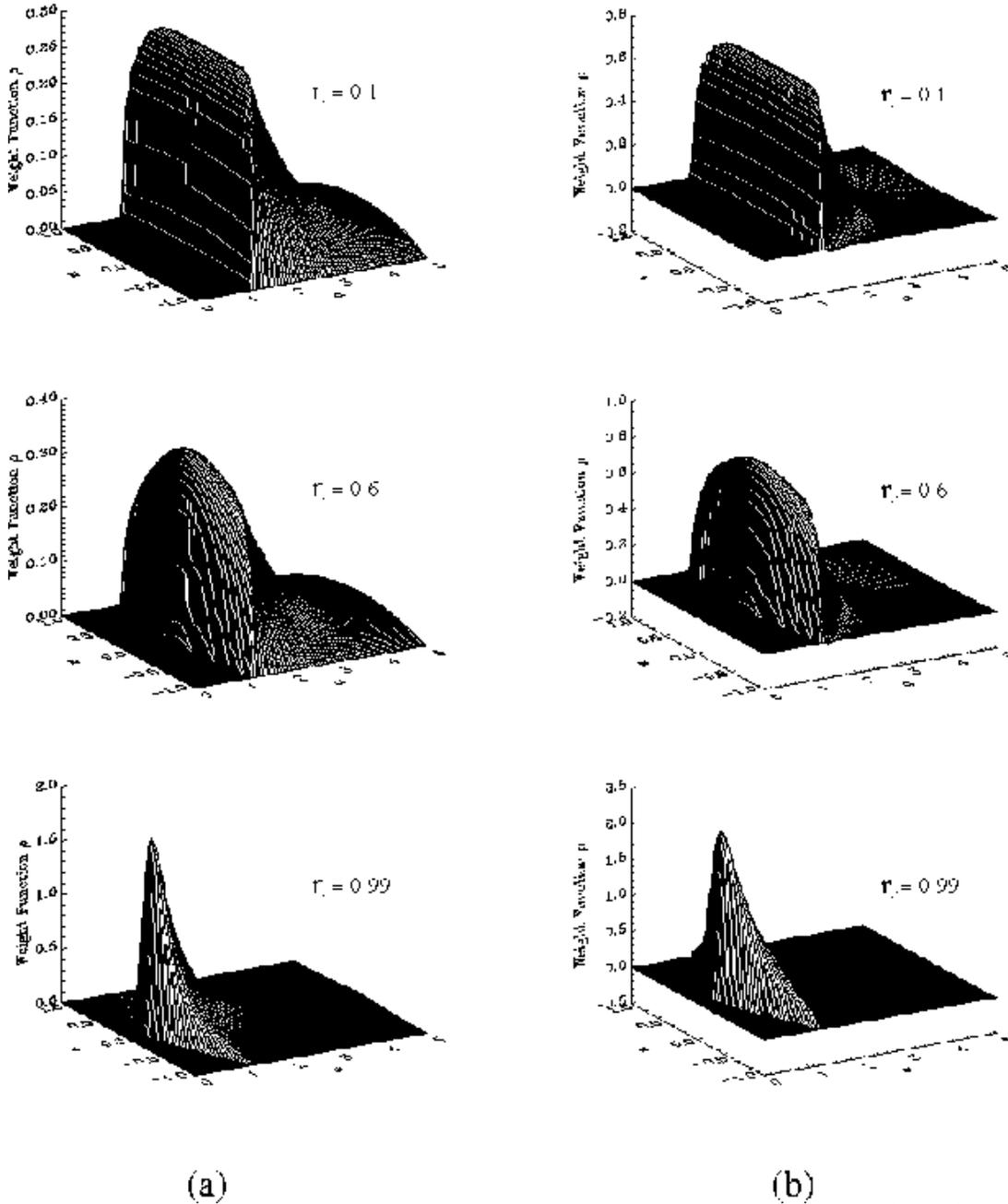,height=18cm} }
\protect\parbox{130mm}{\caption{
Solutions for the bound-state vertex weight function for the $s$-wave
($\ell=0$) and $p$-wave ($\ell=1$) cases are given in (a) and (b) respectively
for a variety of values for the fraction of binding 
$\eta\equiv\protect{\sqrt{P^2}}/2m=M/2m$. The exchange particle ($\sigma$)
mass is $m_\sigma=m/2$ in these solutions. It is convenient to plot the
rescaled weight function 
$\rho(\alpha,z)\equiv\rho_2^{[\ell]}(\alpha,z)/\alpha^2$ in these figures.
}
\protect{\label{fig_pure_ladder}}
}
\end{figure}

\begin{figure}[hbt]
\centering{\
     \psfig{angle=0,figure=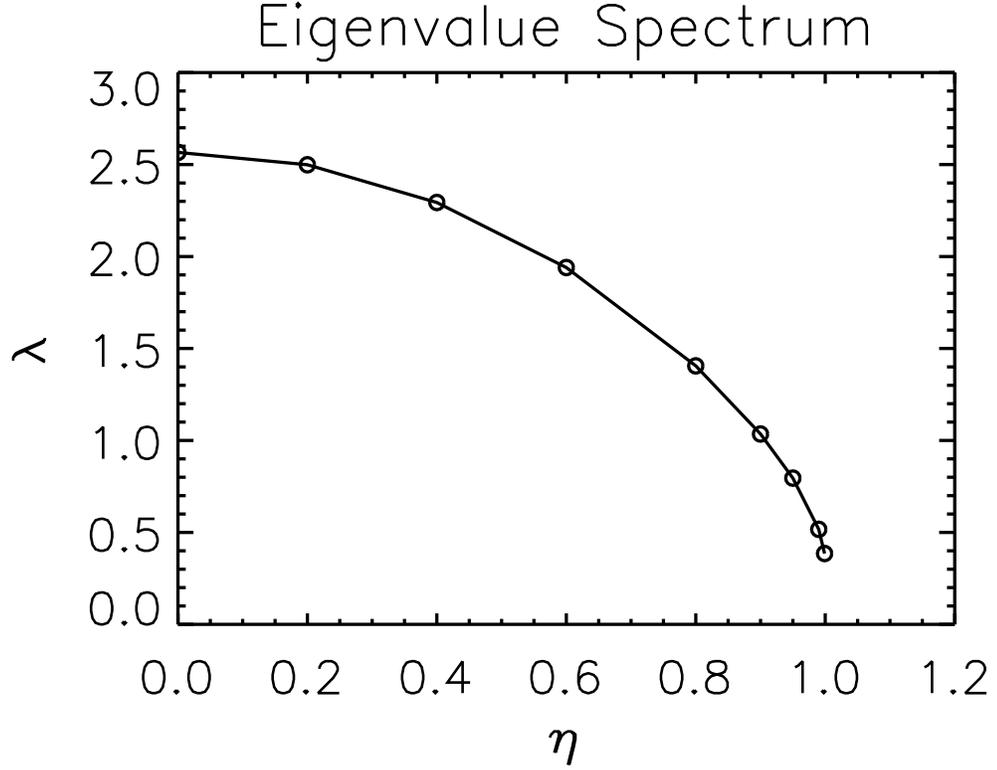,height=11cm} }
\protect\parbox{130mm}{\caption{The $s$-wave ({\it i.e.}, $\ell=0$)
bound-state spectrum for the ladder kernel, where $\lambda\equiv g^2/(4\pi)^2$
and $\eta\equiv\protect{\sqrt{P^2}}/2m=M/2m$. The exchanged particle
($\sigma$) mass is half the constituent mass, $m_\sigma=m/2$, in these
results. These results correspond to those in 
Table~\protect{\ref{table_pure_ladder}}.
}
\label{fig_ladder_spectrum}
}
\end{figure}

We have solved the vertex BS equation, Eq.(\ref{eqn_wght}), for a number
of bound state masses between $P^2=0$ (Goldstone-like bound state) and
$P^2=4m^2$ (the stability threshold). 
Solutions were obtained for orbital excitations up to $\ell=4$ with
no difficulties.
We plot some examples of our solutions for $\ell=0$ and $\ell=1$
in Fig.~\ref{fig_pure_ladder}, and tabulate our results for the
``eigenvalue'' $\lambda\equiv g^2/(4\pi)^2$ in Table~\ref{table_pure_ladder}.
A plot of the spectrum for $\ell=0$, i.e., of $\lambda$ {\it vs.}
the fraction of binding $\eta\equiv\sqrt{P^2}/2m$, 
is given in Fig.~\ref{fig_ladder_spectrum}.
All solutions presented were obtained using a $\sigma$ mass of $m_\sigma=m/2$. 
We have compared our eigenvalues to those obtained in the Wick-rotated
treatment of Linden and Mitter~\cite{L+M}, and have found agreement
to better than 0.03\% for moderate choices of the ($\alpha,z$)-grid. 
This is an improvement
in accuracy of at least one order of magnitude over the results we have 
obtained previously for the BS amplitude. Furthermore, much greater
accuracy is possible through an increase in the number of grid points
used in the numerical integration should it be desired for whatever reason.

\subsection{Dressed Ladder Kernel}

For the ``dressed ladder'' case, the scattering kernel is given by 
$I=-iK=(ig)D_\sigma(ig)$, 
where $D_\sigma$ is the renormalised $\sigma$-propagator 
at one-loop order. This is simply the sum of the pole term, 
Eq.(\ref{pure_ladder_kernel}), and a continuum part, and is given by
	\begin{equation}
		D_\sigma[(p-q)^2]=\int\limits_{0}^{\infty}ds
		 \frac{\rho_\sigma(s)}{s-(p-q)^2-i\epsilon},
	\label{sigmaprop}
	\end{equation}
where $\rho_\sigma(s)$=$\rho_{\rm pole}(s)+\rho_{\rm cont}(s)$.
$\rho_{\rm pole}(s)$ is simply $\delta(s-m_\sigma^2)$, and
it can be shown that, to one-loop order, $\rho_{\rm cont}(s)$
is given by the following expression:
	\begin{equation}
		\rho_{\rm cont}=\lambda\theta(s-4m^2)
			\sqrt{\frac{s-4m^2}{s}}
			\frac{1}{\Delta(s)}, 
	\end{equation}
where $\Delta(s)$ is the following function:
	\begin{eqnarray}
		\Delta(s)&=&\left[m_\sigma^2-s-\lambda\left\{
			2\sqrt{\frac{4m^2-m_\sigma^2}{m_\sigma^2}}
			\arctan\sqrt{\frac{m_\sigma^2}{4m^2-m_\sigma^2}} 
			\right.\right. \nonumber \\
			&&\qquad+\left.\left.
			\frac{4(m_\sigma^2-s)}{m_\sigma^2}\left(
			\frac{m^2}{\sqrt{m_\sigma^2(4m^2-m_\sigma^2)}}
			\arctan\sqrt{\frac{m_\sigma^2}{4m^2-m_\sigma^2}}
			-\frac{1}{4}\right)\right.\right. \nonumber \\
			&&\qquad\qquad+\left.\left.
			2\sqrt{\frac{s-4m^2}{s}}\ln\left(
				\frac{\sqrt{s}+\sqrt{s-4m^2}}{2}
				\right)
			\right\}
			\right]^2 \nonumber \\
		&+& \lambda^2\pi^2\left(\frac{s-4m^2}{s}\right).
	\end{eqnarray}

Note that the use of Eq. (\ref{sigmaprop}) introduces an extra integration
(over the mass parameter $s$). We performed this numerically using gaussian
quadrature; 10 to 15 quadrature points in $s$ provide solutions of 
satisfactory accuracy.

We have solved Eq. (\ref{eqn_wght}) for the dressed ladder kernel above,
with the pole being situated at $m_\sigma=m$, for various values of
the bound state mass squared, $P^2$. 
As for the pure ladder case above, solutions have been obtained
up to $\ell=4$.  For example, our $s$-wave ($\ell=0$) eigenvalue for
$P^2=3.24m^2$ and for the exchange particle pole at $m_\sigma=m$
is $\lambda=1.516$ for a grid of $80 \times 41$ and is
$\lambda=1.518$ for a grid of $150 \times 91$. The corresponding
Linden and Mitter value is $\lambda_E=1.518$. Even with a relatively
coarse grid high accuracies result.  Similarly, we
have found for other values of $P^2$ that an accuracy of 0.3\% or better
is routinely attained, even with the use of the coarse $80 \times 41$ grid.
As the above results demonstrate higher accuracy is easily obtained at the
cost of more CPU time.

\begin{figure}[hbt]
\centering{\
     \psfig{angle=0,figure=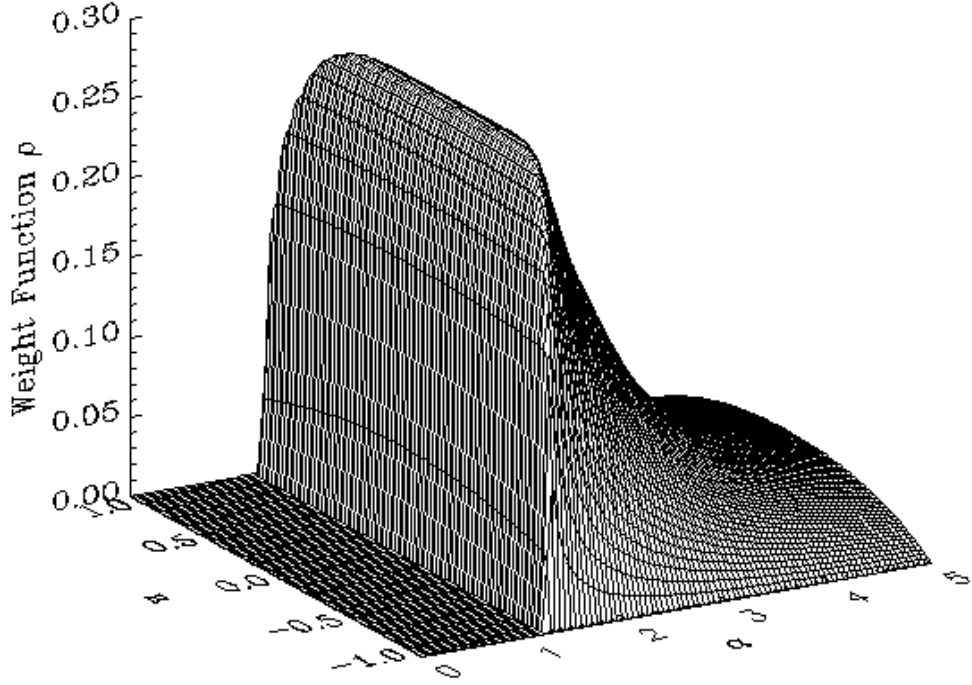,height=11.7cm} }
\protect\parbox{130mm}{\caption{Rescaled weight function 
$\rho(\alpha,z)\equiv\rho_2^{[0]}(\alpha,z)/\alpha^2$ for the 
dressed ladder kernel with pole at $m_\sigma^2=m^2$, 
and a bound-state mass of $\eta\equiv\protect{\sqrt{P^2}}/2m=0.1$.} 
\label{fig_self_energy}
}
\end{figure}

We plot the re-scaled weight function, $\rho_2^{[0]}/\alpha^2$, for the
case $P^2=0.04m^2$ in Fig.~\ref{fig_self_energy} for purposes of
comparison with the corresponding ladder case. We have chosen this 
value of $P^2$ since a smaller bound-state mass corresponds to tighter
binding and hence larger coupling, which should enhance the effect of this
one-loop self-energy insertion. We see that the shape of the weight function
is not qualitatively very different from the ladder case.

\subsection{Generalised Kernel}

This particular example of a ``generalised'' kernel 
is an instance of a scattering kernel
for which Euclidean space solution is not possible. We have solved the
BS equation for this case, in particular for a sum of the pure ladder kernel
as described above and two randomly chosen non-ladder
terms, each with a weight of
$0.25\lambda$, and with the following fixed parameter sets 
(in the $st$ channel):
\begin{eqnarray}
&&\{\gamma,a_{st},b_{st},c_{st},d_{st},e_{st},f_{st}\}= \nonumber \\
&(1)& \{2.25m^2, 0.47261150181, -0.29743163287, 0.58277042955,
    \nonumber \\
&& 0.28282145969,-0.23965580016, 0.32196629047\} \nonumber \\
&(2)& \{2.25m^2, 0.47261150181, 0.29743163287, 0.58277042955,
    \nonumber \\
&&0.28282145969, -0.23965580016, -0.32196629047\}.
\end{eqnarray}
The parameters $\{a_{st},\ldots,f_{st}\}$ were obtained from a set of
values $\{\xi_1,\ldots,\xi_6\}$ produced by a random number
generator (see Appendix~\ref{apdx_a}).
This was done to emphasise that our technique 
produces well-behaved solutions for an {\it arbitrary} kernel.
We have had little difficulty obtaining noise-free solutions for this kernel
for orbital excitations up to $\ell=4$. 

\begin{figure}[hbt]
\centering{\
     \psfig{angle=0,figure=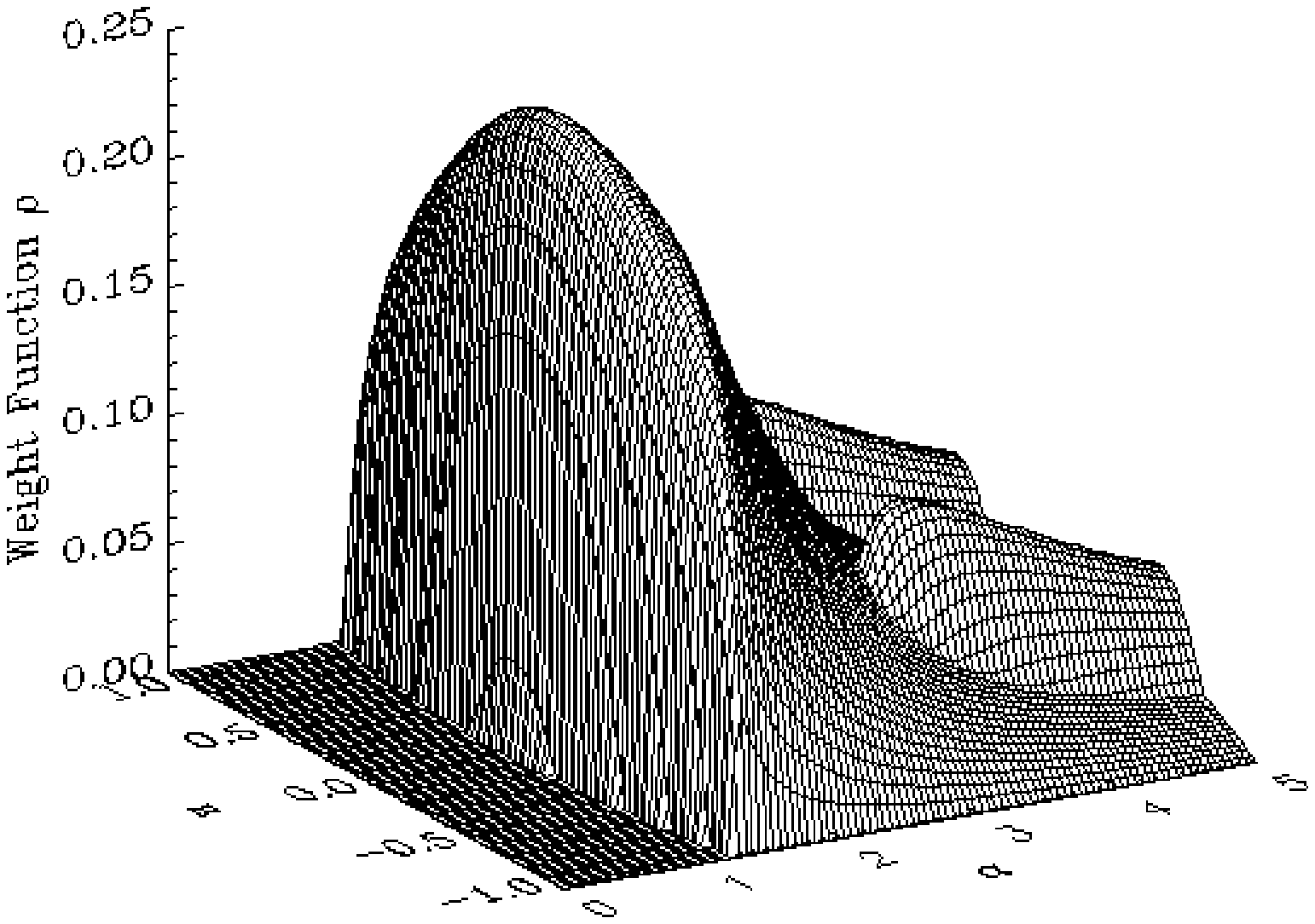,height=11.7cm} }
\protect\parbox{130mm}{\caption{Rescaled weight function 
$\rho(\alpha,z)\equiv\rho_2^{[0]}(\alpha,z)/\alpha^2$ for a random
choice of generalised kernel as described in the text.} 
\label{fig_gnrl_ladder}
}
\end{figure}

This kernel yields an $s$-wave eigenvalue of
$\lambda\equiv g^2/(4\pi)^2=1.3569$ for $P^2=1.44m^2$,
c.f. the ladder value of $\lambda=1.9402$
for the same bound state mass. We therefore
find, as before~\cite{BS_amplitude}, that the additions to the pure ladder 
kernel have enhanced the binding, {\it i.e.} they are attractive.
This is of course to be expected in a scalar theory and was also
observed in Refs.~\cite{Tjon1,Tjon2}. 
Not only is the eigenvalue lower, but additional structure is present
in the vertex weight function (see Fig.~\ref{fig_gnrl_ladder}). In contrast
to the solutions obtained for the BS amplitude in the previous 
work~\cite{BS_amplitude}, we find that there is no
numerical noise in the vertex weight function. If one compares 
the general kernel example solution in Fig.~\ref{fig_gnrl_ladder} 
with the ladder solution for the same bound-state
mass (see Fig.~\ref{fig_pure_ladder}), 
it is readily apparent that there is some additional structure
superimposed upon the weight function, due to the addition of the 
generalised kernel terms.

\section{Summary and Conclusions}
\label{Conclusions}

We have derived a real integral equation for the weight function of the
scalar-scalar Bethe-Salpeter (BS) vertex from the BS equation for
scalar theories without derivative coupling.
This was achieved using the perturbation theory integral representation
(PTIR), which is an extension of the spectral representation for two-point
Green's functions, for both the scattering kernel [Eq.~(\ref{krnl_PTIR})]
and the BS vertex itself [Eq.~(\ref{ptlw_BS_PTIR2})].
The uniqueness theorem of the PTIR and the appropriate application
of Feynman parametrisation then led to the central result of the paper
given in Eq.~(\ref{eqn_wght}).

We have demonstrated that Eq.(\ref{eqn_wght}) is numerically tractable
for several simple kernels, including a randomly chosen case 
where it is not possible to
write the kernel as a sum of ordinary Feynman diagrams. 
Our results for both the pure and dressed
ladder kernels are in excellent agreement with the results obtained
previously in the Wick-rotated approach. The agreement for the pure ladder 
kernel is even better than in Ref.~\cite{BS_amplitude}, vindicating our
decision to solve here the vertex equation rather than the amplitude equation.
We obtained an accuracy in all of our results of approximately 1 in 10$^4$
with modest ($\alpha,z$)-grid choices on a workstation. This can be improved
by using finer grids and larger computers as desired.

Further applications of our formalism are currently being investigated,
particularly the crossed ladder and separable kernels. 
These will not only provide yet another
test of our implementation of the method, as Euclidean space results
are also available for these cases, but also will provide us with an 
opportunity to solve a problem featuring more realistic scattering
kernels.

It is also important to consider
how the PTIR can be extended to include fermions and derivative
coupling, so that we have a covariant framework within which to study,
for example, mesons in QCD using a coupled Bethe-Salpeter---Dyson-Schwinger
equation approach. This would require us to incorporate confinement into
the PTIR, which at this stage remains another important and interesting
challenge. 

{\bf [Those interested in applications of this technique may
request a copy of the computer code from AGW at the given
e-mail address.]}

\begin{acknowledgments}

We thank Stewart Wright for assistance in generating the 
numerical results and the figures and A.W.\ Thomas for some helpful
comments on the manuscript. We also thank V.~Sauli for a careful
proofreading of the manuscript.
This work was supported by the Australian Research Council, by
Scientific Research grant \#1491 of the Japan Ministry of Education and 
Culture, and also in part by grants of
supercomputer time from the U.S. National Energy Research Supercomputer
Center and the Australian National University
Supercomputer Facility.

\end{acknowledgments}

\appendix

\def\gl{\mathrel{\rlap{\lower3pt\hbox{$<$}}
    \raise3pt\hbox{$>$}}}

\section{Algorithm}
\label{apdx_0}

Here we detail the algorithm used in  
our numerical studies of the integral equation:

\begin{equation}
\frac{1}{\lambda}\frac{\rho_n^{[\ell]}(\bar{\alpha},\bar{z})}{\bar{\alpha}^n}
	=
	\int\limits_{0}^{\infty} d\alpha\int\limits_{-1}^{1}dz \;
	\frac{\rho_n^{[\ell]}(\alpha,z)}{\alpha^n}
	\;{}^{\rm tot}\!{\cal K}_n^{[\ell]}
	(\bar{\alpha},\bar{z};\alpha,z;\lambda).
\label{int_eqn}
\end{equation}
where we have explicitly shown the coupling dependence of 
the kernel function. We have defined here an ``eigenvalue''
$\lambda\equiv g^2/(4\pi)^2$. Our rationale for this is as follows:
for a given scattering kernel the integral equation 
(\ref{int_eqn}) may be solved for the bound state mass $P^2=M^2$.  
However, the dependence of the kernel function 
${}^{\rm tot}\!{\cal K}_n^{[\ell]}(\bar{\alpha},\bar{z};\alpha,z)$ 
on the bound state mass $P^{2}$ is highly non--linear and complicated.  
It is therefore convenient and traditional to instead solve the equation 
for the coupling $g^2$, which appears in the weight function 
$\rho_{\rm ch}(\gamma,\vec\xi)$ for the scattering kernel, with 
a fixed bound state mass $P^{2}$.  We first fix the bound state mass $P^2$ 
and regard the integral equation (\ref{int_eqn}) as an ``eigenvalue'' 
problem.  The ``eigenvalue'' is then introduced by factorising the
coupling constant $g^2$ from the scattering kernel weight function 
$\rho_{\rm ch}(\gamma,\vec\xi)$. In this convention, the kernel function
${}^{\rm tot}\!{\cal K}_n^{[\ell]}(\bar{\alpha},\bar{z};\alpha,z;\lambda)$
becomes a power series in $\lambda$ starting from ${\cal O}(1)$ for a 
perturbative scattering kernel.

 Strictly speaking, the integral equation 
(\ref{int_eqn}) is not an ``eigenvalue'' equation, since the kernel 
function itself contains $\lambda$ in the general ({\it i.e.}, non-ladder)
case.  
We thus solve the equation by iteration rather than applying 
methods for eigensystems.  
With an appropriate initial guess for the 
weight function $\rho_n^{[\ell]}(\alpha,z)$ for the BS vertex 
and the coupling constant $\lambda$ we generate the new weight function 
by evaluating the RHS of the integral equation (\ref{int_eqn}).  
The ``eigenvalue'' $\lambda$ associated with the weight function 
is extracted by imposing an appropriate normalization 
condition which we will discuss later.  
This generated weight function and its ``eigenvalue'' are used as inputs
and we obtain updated values, which ought to be closer to the solution 
than the input values, by evaluating the integral.  
We repeat this cycle until both the 
``eigenvalue'' and the weight function converge.  

The normalization condition for the BS vertex or equivalent BS amplitude 
in momentum space is well known and involves the derivative of the 
scattering kernel with respect to the bound state 4-momentum $P$ 
(See Eq.(\ref{normalization}) in Sec.\ref{Representation}).  
When the scattering kernel depends on the total momentum $P$, 
the normalization condition is the integral over two relative momenta; 
one for the BS vertex and the other for the conjugate one.  
The corresponding normalization condition in PTIR form is written as 
the 4-dimensional integral over spectral parameters $\alpha$ and $z$.  
Imposing a condition that involves such a multi--dimensional integral in the 
iteration cycle makes the calculation less accurate and time consuming.  
We shall rather use a suitable normalization condition for the 
weight function $\rho_n^{[\ell]}(\alpha,z)$ during the iteration. 
The physical normalization condition (\ref{normalization}) may be 
imposed by appropriately rescaling the obtained solution. Of course,
the value of $\lambda\equiv g^2/(4\pi)^2$ is unaffected by the choice of
normalisation of the vertex weight function.

Since we are considering bound states whose constituents are of equal  
mass $m$, we expect that a physically reasonable scattering kernel 
$I(p,q;P)$ will give a kernel function 
${\cal K}_n^{[\ell]}(\bar{\alpha},\bar{z};\alpha,z;\lambda)$ 
symmetric under the transformation $\bar z \rightarrow -\bar z$.  
The weight function $\rho_n^{[\ell]}(\alpha,z)$ is then either 
symmetric or anti--symmetric in $z \rightarrow - z$.  
For a symmetric solution the following normalization is convenient:

\begin{equation}
\int\limits_{0}^{\infty} d\alpha\int\limits_{-1}^{1}dz \;
	\frac{\rho_n^{[\ell]}(\alpha,z)}{\alpha^n} = 1,
\label{sym_norm}
\end{equation}

\noindent
provided that the integral does not identically vanish.  
From Eq.~(\ref{tot_kernel}), the kernel function is given by the difference
of two terms, {\it viz}
%${}^{\rm tot}\!{\cal K}_n^{[\ell]}(\bar{\alpha},\bar{z};\alpha,z;\lambda)$
${\cal K}_n^{[\ell]}(\bar{\alpha},\bar{z};0,0;\lambda)-
{\cal K}_n^{[\ell]}(\bar{\alpha},\bar{z};\alpha,z;\lambda)$.  
With the normalization condition (\ref{sym_norm}) the integral equation 
(\ref{int_eqn}) can be written as an inhomogeneous one:

\begin{equation}
\frac{1}{\lambda}\;\frac{\rho_n^{[\ell]}(\bar{\alpha},\bar{z})}
	{\bar{\alpha}^n}=
	{\cal K}_n^{[\ell]}(\bar{\alpha},\bar{z};0,0;\lambda)
	-\int\limits_{0}^{\infty} d\alpha\int\limits_{-1}^{1}dz \;
	\frac{\rho_n^{[\ell]}(\alpha,z)}{\alpha^n}\;{\cal K}_n^{[\ell]}
	(\bar{\alpha},\bar{z};\alpha,z;\lambda).
\label{normal_eqn}
\end{equation}

\noindent
On the other hand, the integral of the weight function over $\alpha$ and $z$ 
for an anti--symmetric solution vanishes identically.  
Then the integral equation (\ref{int_eqn}) is written as

\begin{equation}
\frac{1}{\lambda}\frac{\rho_n^{[\ell]}(\bar{\alpha},\bar{z})}
	{\bar{\alpha}^n}=
	-\int\limits_{0}^{\infty} d\alpha\int\limits_{-1}^{1}dz \;
	\frac{\rho_n^{[\ell]}(\alpha,z)}{\alpha^n}\;{\cal K}_n^{[\ell]}
	(\bar{\alpha},\bar{z};\alpha,z;\lambda).
\label{abnormal_eqn}
\end{equation}

\noindent
As discussed in the main text, 
the bound state whose BS amplitude $\Phi(p,P)$ and 
equivalent vertex $\Gamma(p,P)$ are anti--symmetric 
under the transformation $P\cdot p \rightarrow - P\cdot p$ 
with fixed $p^{2}$ and $P^{2}$ has a negative norm and is called a 
``ghost''\cite{Nakanishi_survey}.  This symmetry corresponds to 
the one: $\bar z \rightarrow -\bar z$ in PTIR form.  Thus a bound state 
whose weight function is anti--symmetric in $z$--reflection and which  
satisfies the homogeneous integral equation (\ref{abnormal_eqn}) is 
a ``ghost'' state.  
We hereafter concentrate on the ``normal'' solutions, namely 
$z$--symmetric ones, and on the integral equation 
(\ref{normal_eqn}).  

It can be shown that the inhomogeneous term vanish unless 
$\bar\alpha > \alpha_{\rm th}$, where $\alpha_{\rm th}$ is the threshold 
of the weight function depending on the value of $\bar z$ 
for a given  scattering kernel.  
For a one--$\sigma$ exchange kernel with the mass $\mu$, 
the threshold can be written as 

\begin{equation}
	\alpha_{\rm th}(\bar z)=
		\left(\left(m^2-(1-\bar z^2)\frac{P^2}{4}\right)^{1/2}
			+\mu \right)^2
		-\left(m^2 - (1-\bar z^2) \frac{P^2}{4}\right)
	\label{lad_th}
\end{equation}

\noindent
This threshold determines the support of the weight function $\rho$ or 
equivalently $\varphi$ for the normal solution.  
Although we cannot write the threshold in a simple form like 
(\ref{lad_th}) for general scattering kernels, we can extract it 
numerically by analyzing the inhomogeneous term.  
On the other hand, the kernel function 
${\cal K}_n^{[\ell]}(\bar{\alpha},\bar{z};\alpha,z;\lambda)$ 
has the support property for a given $\bar\alpha$, $\bar z$ and $z$ 
that it vanishes unless $\alpha$ is less than some value $\alpha_{\rm max}$. 
For the case of a one--$\sigma$ exchange kernel it is given by
 
\begin{eqnarray}
	\alpha_{\rm max}(\bar\alpha,\bar z,z)&=&\left(
		\left( \bar\alpha + 
			\left(m^2-(1-\bar z^2)\frac{P^2}{4}\right)^{1/2}
		- \mu \right)^2
		-\left(m^2 - (1-\bar z^2) \frac{P^2}{4}\right)
		\right) \frac{1\mp z}{1\mp\bar z}, \\ \nonumber
		& &\qquad\qquad\qquad\qquad\qquad\qquad\qquad
		\qquad\qquad\qquad\qquad\qquad\qquad\qquad
		 \hbox{for }\bar z \gl z.
	\label{lad_upth}
\end{eqnarray}

\noindent
As in the case of $\alpha_{\rm min}$ the analytic form of the upper 
limit $\alpha_{\rm max}$ 
is unknown, so we extract the corresponding upper limit 
numerically for general scattering kernels.  
Writing these limits of the integral explicitly 
the integral equation we use is then
\begin{equation}
\frac{1}{\lambda}\;\frac{\rho_n^{[\ell]}(\bar{\alpha},\bar{z})}
	{\bar{\alpha}^n}=
	{\cal K}_n^{[\ell]}(\bar{\alpha},\bar{z};0,0;\lambda)
	-\int\limits_{-1}^{1}dz \int_{\alpha_{\rm th}(z)}
		^{\alpha_{\rm max}(\bar\alpha,\bar z,z)} d\alpha\;
	\frac{\rho_n^{[\ell]}(\alpha,z)}{\alpha^n}
	\;{\cal K}_n^{[\ell]}
	(\bar{\alpha},\bar{z};\alpha,z;\lambda).
\label{eqntosolve}
\end{equation}

We evaluate the integral over $\alpha$ and $z$ in the RHS of 
Eq.~(\ref{eqntosolve}) as follows.  
Recall that the kernel function 
${\cal K}_n^{[\ell]}(\bar{\alpha},\bar{z};\alpha,z;\lambda)$ is given 
by the following integral:
\begin{equation}
	{\cal K}_n^{[\ell]}(\bar{\alpha},\bar{z};\alpha,z;\lambda) \equiv 
	\sum_{\rm ch}\int\limits_\Omega d\vec{\xi}\int\limits_0^\infty d\gamma\;
	\left[\frac{1}{g^2}\rho_{\rm ch}(\gamma,\vec{\xi})\right]
	K_n^{[\ell]}(\bar{\alpha},
	\bar{z};\alpha,z;\gamma,\vec{\xi})\;.
	\label{krnl}
\end{equation}
We thus start by replacing the integrations over Feynman parameters $\vec\xi$ 
and the spectral variable $\gamma$ by summations over discretised 
variables.  %(]
Secondly, we map the semidefinite range of $\alpha \in [0, \infty)$ to the 
finite one $y \in [0,1]$:
\begin{equation}
	\alpha=\alpha_0 + C\frac{y}{1-y},
\end{equation}
where $\alpha_0$ and $C$ are some constants which should be chosen 
such that the weight function is largest around the mapped 
variable $y\sim 1/2$.  We then discretise both $y$ (equivalently $\alpha$) 
and $z$ and prepare the initial weight function on this grid.  
For each cycle of the iteration we perform the integral as follows.  
We first evaluate the $\alpha$ integral for a given point in the 
$\bar\alpha$ and 
$\bar z$ plane and on the $z$ grid.  
For each value of the discretised $\vec\xi$ and $\gamma$ we extract 
the support of the the kernel function 
$K_n^{[\ell]}(\bar{\alpha},\bar{z};\alpha,z;\lambda)$.  
We then divide the integral range 
$[\alpha_{\rm th}(z), \alpha_{\rm max}(\bar\alpha,\bar z,z)]$ into 
subranges according to the support of the kernel function with 
discretised $\vec\xi$ and $\gamma$.  As discussed in Appendix~\ref{Singularity} 
the kernel function 
$K_n^{[\ell]}(\bar{\alpha},\bar{z};\alpha,z;\lambda)$ 
may diverge as an integrable 
square root singularity at the boundary of the support.  
While this is always the case for the one $\sigma$ exchange kernel, 
the kernel may take a finite value in general.  We thus choose 
an appropriate integration method to perform 
the integration over $\alpha$ for each subrange.  
The weight function at arbitrary $\alpha$ is evaluated 
by interpolating the values of $\rho$ on the grid.  
We perform the $\alpha$ integral in this way for each grid point of $z$ 
and the integral over $z$ is performed by interpolating these 
values.  With this careful treatment of integrable square root singularities 
we need not introduce any regularization or cutoff parameters.  
Furthermore, this method allows us to choose the $\bar\alpha$ and $\bar z$ 
grid for the newly generated weight function independent of 
the $\alpha$ and $z$ grid. We optimise the ``new'' grid by analyzing 
the shape of the ``old'' weight function used 
in the RHS of Eq.(\ref{eqntosolve}).  
The eigenvalue is evaluated using the normalization condition (\ref{sym_norm}). 
 
\newcommand{\nc}{\newcommand}
\nc{\be}{\begin{equation}}
\nc{\ee}{\end{equation}}
\nc{\bea}{\begin{eqnarray}}
\nc{\eea}{\end{eqnarray}}
\nc{\beas}{\begin{eqnarray*}}
\nc{\eeas}{\end{eqnarray*}}
\nc{\half}{{\textstyle \frac{1}{2}}}
\nc{\qtr}{{\textstyle \frac{1}{4}}}
\nc{\halfP}{\frac{P}{2}}
\nc{\qtrP}{\frac{P^2}{4}}
\nc{\ul}{\underline}
\nc{\non}{\nonumber}
\nc{\noi}{\noindent}

\nc{\Kint}{\frac{\rho(\gamma,\vec{\xi})}{\gamma-(aq^2+ bp\cdot q + cp^2
   +dP^2 +eq\cdot P + fp\cdot P) - i\epsilon}}
\nc{\solidq}{{\cal Y}_\ell^{\ell_z}(\Lambda^{-1}(P)q)}
\nc{\solidp}{{\cal Y}_\ell^{\ell_z}(\Lambda^{-1}(P)p)}
\nc{\loopint}{\int\limits\frac{d^4q}{(2\pi)^4 i}}
\nc{\vertwt}{\rho_n^{[\ell]}(\alpha,z)}

\section{PTIR for Scattering Kernel}
\label{apdx_a}

In this appendix we list the dimensionless coefficients
$\{a_{\rm ch},b_{\rm ch},c_{\rm ch},\dots,f_{\rm ch}\}$
in Eq. (\ref{krnl_PTIR_sum})
for different channels $\{{\rm ch}\}=\{{st}\},\{{tu}\},\{{us}\}$
in terms of the Feynman parameters $\xi_i$ defined in Eq. (\ref{krnl_PTIR}).

\begin{center}
\begin{tabular}{cccc}
	\hline
	 & $st$	& $tu$ & $us$  \\
	\hline\hline
	$\quad a_{\rm ch}\quad$
		& $\xi_1+\xi_2+\xi_6$
		& $\xi_1+\xi_2+\xi_5+\xi_6$	& $\xi_1+\xi_2+\xi_5$  \\
	$b_{\rm	ch}$ & $-2 \xi_6$ &	$2(\xi_6-\xi_5)$ & $2\xi_5$	 \\
	$c_{\rm	ch}$
		& $\xi_3+\xi_4+\xi_6$
		& $\xi_3+\xi_4+\xi_5+\xi_6$	& $\xi_3+\xi_4+\xi_5$  \\
	$d_{\rm	ch}$ & $\quad{1\over 4}(\xi_1+\xi_2+\xi_3+\xi_4)+\xi_5\quad$
		& $\quad{1\over	4}(\xi_1+\xi_2+\xi_3+\xi_4)\quad$
		& $\quad{1\over	4}(\xi_1+\xi_2+\xi_3+\xi_4)+\xi_6\quad$	 \\
	$e_{\rm	ch}$ & $\xi_1-\xi_2$ & $\xi_1-\xi_2$ & $\xi_1-\xi_2$  \\
	$f_{\rm	ch}$ & $\xi_3-\xi_4$ & $\xi_3-\xi_4$ & $\xi_3-\xi_4$  \\
	\hline
\end{tabular}\label{table_abc}
\end{center}

Beginning with the above definitions for the scattering kernel parameters
in terms of the Feynman parameters $\vec{\xi}$,
and noting that $\sum_{i=1}^6 \xi_i=1$, it is possible to prove the following
relations between the kernel parameters, for all three channels:
\begin{eqnarray}
ac-\frac{b^2}{4} &\geq& 0                                \non \\
\left|f\pm\frac{b}{2}\right| &\leq& c 			\non \\
\left|af-\frac{eb}{2}\right| &\leq& ac-\frac{b^2}{4}.
\label{skpara_relns}
\end{eqnarray}

\section{Kernel Function}
\label{apdx_b}

In this appendix we detail our derivation of the real integral equation
for the BS vertex. We begin with the following PTIR form of the 
bound-state vertex~\cite{Nakanishi63}:
\be
\Gamma^{[\ell,\ell_z]}(q,P)=\solidq\int\limits_{0}^{\infty}d\alpha\int\limits_{-1}^{1}dz
\frac{\rho_n^{[\ell]}(\alpha,z)}{[F(\alpha,z;q,P)-i\epsilon]^n}.
\label{basicvertex}
\ee
In Eq.~(\ref{basicvertex}), $\solidq$ is the solid harmonic for a bound
state with angular momentum quantum numbers $\ell$ and $\ell_z$, $\vertwt$
is the PTIR weight function for the bound state vertex function, and $n$ is a 
dummy parameter. The
function $F$ is given by

\bea
F(\alpha,z;q,P)&=&\alpha+\frac{1+z}{2}(m^2-(q+\half P)^2)+
    \frac{1-z}{2}(m^2-(-q+\half P)^2)\nonumber\\
               &=&\alpha+m^2-(q^2+zq\cdot P+\qtr P^2).
\label{f-fn}
\eea

We proceed by substituting this form of the vertex into the vertex BSE,
Eq.~(\ref{BSE_Gamma}), and combining the various factors on the right 
hand side of
the resultant equation (bare propagators, scattering kernel and vertex
PTIR) using Feynman parametrisation.
We first combine the bare
propagators for the scalar constituents with the denominator of the
vertex
PTIR:

\bea
& &\hspace{-1.6cm}D(q+\half P)D(-q+\half P)\frac{1}{[F(\alpha,z;q,P)-i\epsilon]^n}\nonumber\\
&=&\frac{1}{2} \frac{\Gamma(n+2)}{\Gamma(n)\Gamma(2)}\int\limits_{-1}^{1}d\eta
        \int\limits_{0}^{1}dt\;t^{n-1}(1-t)\frac{1}{[F(t\alpha,tz+(1-t)\eta;
        q,P)-i\epsilon]^{n+2}}.
\label{prop_with_vertex}
\eea
        
We now combine the factor $1/[F(\ldots)]^{n+2}$ from the integrand with
the denominator of the PTIR for the scattering kernel 
(see Eq. (\ref{krnl_PTIR_sum})). 
After some algebra this yields

%\bea
%& &\hspace{-1.6cm}\frac{1}{\gamma-(aq^2+bp\cdot q+cp^2+dP^2+eq\cdot P+fp\cdot P)-i\epsilon}
%\frac{1}{[F(t\alpha,tz+(1-t)\eta;q,P)-i\epsilon]^{n+2}}\non \\
%&=&\frac{\Gamma(n+3)}{\Gamma(1)\Gamma(n+2)}\int\limits_{0}^{1}dx\;\frac{x^{n+1}}
%{(1-x)^{n+3}}\frac{1}{(y+a)^{n+3}} \non \\
%&\times&\frac{1}{\left[\frac{c(y+a)-\frac{b^2}{4}}{y+a} F({\cal A,Z};p,P)-
%\left(q+\frac{bp+(e+zy)P}{2(a+y)}\right)^2-i\epsilon\right]^{n+3}},
%\label{kernel_with_prop/vert}
%\eea
%
\bea
& &\hspace{-1.6cm}\frac{1}{\gamma-(aq^2+bp\cdot q+cp^2+dP^2+eq\cdot P+fp\cdot P)-i\epsilon}
\frac{1}{[F(t\alpha,tz+(1-t)\eta;q,P)-i\epsilon]^{n+2}}\non \\
&=&\frac{\Gamma(n+3)}{\Gamma(1)\Gamma(n+2)}\int\limits_{0}^{1}dx\;\frac{x^{n+1}}
{(1-x)^{n+3}}\frac{1}{(y+a)^{n+3}} \non \\
&\times&\frac{1}{[\{[c(y+a)-(b^2/4)]/(y+a)^2 \}F'-
\{q+[bp+(e+[tz+(1-t)\eta]z)P]/2(a+y)\}^2-i\epsilon]^{n+3}},
\label{kernel_with_prop/vert}
\eea
where

\bea
F'&\equiv&F[{\cal A}(t\alpha,tz+(1-t)\eta;y),{\cal Z}(tz+(1-t)\eta;y);p,P],
	\non \\
{\cal A}(\alpha,z;y)&\equiv&\frac{1}{c(y+a)-\frac{b^2}{4}}
	\Biggl[\left\{\alpha+m^2-(1-z^2)\frac{P^2}{4}\right\}(y+a)^2 
 	\non \\
&+&\left\{\gamma-(a+c)m^2-a\alpha+(a+c-4d+2z(e-az))\frac{P^2}{4}\right\}
	(y+a)\non \\
&+&\frac{b^2}{4}\left\{m^2-\left(1-\left(\frac{-2(az-e)}{b}\right)^2
	\right)\frac{P^2}{4}\right\}\Biggr] \non \\
{\cal Z}(z;y)&\equiv&\frac{f(y+a)-\frac{b}{2}(e+yz)}{c(y+a)-
	\frac{b^2}{4}} \non \\
y&\equiv&\frac{x}{1-x}.
\label{AandZdefs}
\eea

Note that the parameters $\{a,\ldots,f\}$ do not explicitly have the 
subscript ``ch'' attached to them in this instance (cf. Appendix~\ref{apdx_a}
 and
Eq. (\ref{krnl_PTIR})), for the sake of brevity. We will also, for the time 
being, omit for brevity
the sum over channels $st$, $tu$, $us$ in the kernel function.

Having combined all factors on the right hand side, we are now in a position
to perform the integral over the loop momentum $q$. To do so, we must utilise 
the following property of the solid harmonics, 

\begin{equation}
	\int\limits d^3q {\cal F}(\vec q\,^2)
		{\cal Y}_\ell^{\ell_z}(\vec q+\vec p)
	={\cal Y}_\ell^{\ell_z}(\vec p)\int\limits d^3q
		{\cal F}(\vec q\,^2),
  \label{solid_harm_prop}
\end{equation}
where $F$ is a sufficiently rapidly decreasing function which
gives a finite integral~\cite{Nakanishi63}.
Also note that $\Lambda^{-1}(P)$ boosts the 4-vector $P$ to rest, {\em i.e.}
$\Lambda^{-1}(P)P=(\sqrt{P^2},\vec{0})$, and that the solid harmonics are
functions purely of the three-vector part of their argument, {\em i.e.}
${\cal Y}_\ell^{\ell_z}(p)=|\vec{p}\/|^\ell Y_\ell^{\ell_z}(\hat{p})$ (here 
$\hat{p}=\vec{p}/|\vec{p}\/|$). Bearing this in mind, we obtain for 
the loop integral

\bea
& &\hspace{-2cm}\int\limits\frac{d^4q}{(2\pi)^4i}\frac{\solidq}
	{\left[\widetilde{M}^2-\left(q+\frac{bp+(e+zy)P}{2(a+y)}\right)^2-
	i\epsilon\right]^{n+3}}\non \\
&=&\left(-\frac{b}{2}\right)^{\ell}\frac{1}{(a+y)^\ell}\frac{\Gamma(n+1)}
	{(4\pi)^2\Gamma(n+3)}\frac{\solidp}{(\widetilde{M}^2-
	i\epsilon)^{n+1}},
\eea 

\noi where $\widetilde{M}^2$ is simply that part of the denominator of
Eq.~(\ref{kernel_with_prop/vert}) that does not depend at all on the 
4-momentum $q$.

Ignoring integrations over weight functions for the moment, we have, after
performing the loop momentum integral, the following result:

\bea
&&\hspace{-2cm}\loopint I(p,q;P)D(q+\half P)D(-q+\half P)
	\frac{\solidq}{[F(\alpha,z;q,P)-i\epsilon]^n}  \non\\
&=&\frac{1}{(4\pi)^2}\frac{1}{2}\left(-\frac{b}{2}\right)^\ell
	\frac{\Gamma(n+1)}{\Gamma(n)}  \non \\
&\times&\hspace{5mm}\int\limits_{-1}^{1}d\eta\int\limits_{0}^{1}dt\;t^{n-1}(1-t)
	\int\limits_0^1 dx \frac{1}{(1-x)^2}\frac{y^{n+1}}{(y+a)^\ell}
	\frac{(y+a)^{n-1}}{[c(y+a)-\frac{b^2}{4}]^{n+1}} \non \\
&\times&\hspace{1.5cm}\frac{\solidp}{\left[F({\cal A}(t\alpha,tz+(1-t)\eta;y),
	{\cal Z}(tz+(1-t)\eta;y);p,P)-i\epsilon\right]^{n+1}}.
\label{post-Feynman}
\eea

In order to obtain a real integral equation involving only weight functions,
it is necessary to recast the last factor in Eq.~(\ref{post-Feynman}) 
in a form similar to that found in
the vertex PTIR, Eq.~(\ref{basicvertex}). To proceed we therefore insert 
the trivial integral

\be
\int\limits_{-1}^{1}d\bar{z}\;\delta\left(\bar{z}-{\cal Z}(tz+(1-t)\eta;y)\right)=1
\ee

\noi into the right-hand side of (\ref{post-Feynman}), and eliminate the 
integration over $t$ by rewriting the delta function in terms of $t$.
We are permitted to do this because the function ${\cal Z}$ is bounded between
-1 and 1. That this is true is easily seen by observing that ${\cal Z}$ is
monotonic in the variable $y$, with $y \in [0,\infty)$, and then by taking
the limits $y\rightarrow 0$ and $y \rightarrow \infty$. The former limit gives
\be
{\cal Z}(z;y\rightarrow 0)=\frac{af - \frac{eb}{2}}{ac - \frac{b^2}{4}},
\ee
and so from the third inequality in Eq. (\ref{skpara_relns}) 
we have that $|{\cal Z}(z;y\rightarrow 0)|\leq 1$.
The second limit gives
\be
{\cal Z}(z;y\rightarrow \infty)=\frac{f - \frac{b}{2}z}{c},
\label{limit-infty}
\ee
which allows us to use the second inequality in Eq. (\ref{skpara_relns}) 
to conclude that 
$|{\cal Z}(z;y\rightarrow \infty)| \leq 1$, given that (\ref{limit-infty})
is monotonic in $z$. Since $|{\cal Z}| \leq 1$ in these two limits,
${\cal Z}$ must be bounded between $-1$ and 1 for all $y\in[0,\infty)$.

The insertion of this integral gives us

\bea
&&\frac{1}{(4\pi)^2}\frac{1}{2}\left(-\frac{b}{2}\right)^\ell
	\left|-\frac{2}{b}\right|\frac{\Gamma(n+1)}{\Gamma(n)}
	\int\limits_{-1}^1d\bar{z}\int\limits_0^1\frac{dx}{(1-x)^2}\frac{y^{n+1}}{(y+a)^\ell}
	\frac{(y+a)^{n-1}}{h^{n+1}} \non \\
&\times&\left\{\theta(z-G)\theta(G+1)\int\limits_{-1}^G d\eta\frac{h}{y}
	\frac{z-G}{(z-\eta)^2}\left(\frac{G-\eta}{z-\eta}\right)^{n-1}
	\frac{\solidp}{[F({\cal A}(t_0\alpha,G),\bar{z};p,P)-i\epsilon]^{n+1}}
	\right. \non \\
&&\left.-\theta(G-z)\theta(1-G)\int\limits_G^1 d\eta \frac{h}{y}
	\frac{z-G}{(z-\eta)^2}\left(\frac{G-\eta}{z-\eta}\right)^{n-1}
	\frac{\solidp}{[F({\cal A}(t_0\alpha,G),\bar{z};p,P)-i\epsilon]^{n+1}}
	\right\},
\label{t_integration}
\eea
where we have introduced the following:

\bea
t_0&\equiv&\frac{G-\eta}{z-\eta}\non \\
G&\equiv&G(\bar{z};y)\equiv\frac{1}{y}\left[\frac{-2(c\bar{z}-f)}{b}(y+a)
	-(-\frac{b}{2}\bar{z}+e)\right]\non \\
h&\equiv&h(y)\equiv c(y+a)-\frac{b^2}{4}.
\eea

We next make a change of variable $\eta\rightarrow\bar{\alpha}$, such that

\bea
\bar{\alpha}&=&\frac{y(y+a)}{h}t_0\alpha + \frac{g(\bar{z};y)}{h}, \non
\\
g(\bar{z};y)&\equiv&A'(\bar{z})(y+a)^2+B'(\bar{z})(y+a)+C'(\bar{z})\non \\
A'(\bar{z})&\equiv&m^2-\left(1-\left(\frac{-2(c\bar{z}-f)}{b}\right)^2\right)
	\frac{P^2}{4}\non \\
B'(\bar{z})&\equiv&\gamma-(a+c)m^2+(a+c-4d-2\bar{z}(c\bar{z}-f))\frac{P^2}{4}
	\non \\
C'(\bar{z})&\equiv&\frac{b^2}{4}\left(m^2-(1-\bar{z}^2)\right)\frac{P^2}{4}
\eea

Note that for brevity we sometimes write $g(\bar{z};y)$ as $g$ below.
This should not be confused with the coupling strength $g$, since the
meaning should be clear from the context.

With this
transformation the factor in braces in Eq. (\ref{t_integration})
becomes

\bea
& &\left\{\theta(z-G)\theta(G+1)\int\limits_{g/h}^{R_+\alpha+g/h} d\bar{\alpha}
	\frac{1}{\alpha^n}\frac{h^2}{y}\frac{1}{(y(y+a))^n}(h\bar{\alpha}
	-g)^{n-1}\right.\non \\
& &\left.+\theta(G-z)\theta(1-G)\int\limits_{g/h}^{R_-\alpha+g/h} d\bar{\alpha}
	\frac{1}{\alpha^n}\frac{h^2}{y}\frac{1}{(y(y+a))^n}(h\bar{\alpha}
	-g)^{n-1}\right\} \non \\
&\times&\frac{\solidp}{[F(\bar{\alpha},\bar{z};p,P)-i\epsilon]^{n+1}},
\label{eta_to_alpha}
\eea
with the functions $R_\pm$ being defined by

\be
R_\pm(\bar{z},z;y)\equiv\frac{y(y+a)}{h(y)}\frac{G(\bar{z};y)\pm1}{z\pm1}.
\ee

We may use the following to shift the limits of integration of $\bar{\alpha}$
to $(-\infty,\infty)$:

\be
\int\limits_a^b d\bar{\alpha}=\int\limits_a^\infty d\bar{\alpha} - \int\limits_b^\infty d\bar{\alpha}
	=\int\limits_{-\infty}^\infty d\bar{\alpha}[\theta(\bar{\alpha}-a)
	-\theta(\bar{\alpha}-b)].
\ee

The expression (\ref{eta_to_alpha}) then becomes

\bea
& &\sum_\mp \theta(\pm(G-z))\theta(1\mp G)\int\limits_{-\infty}^\infty d\bar{\alpha}
	\frac{1}{\alpha^n}\frac{h^{n+1}}{y^{n+1}(y+a)^n}\left(\bar{\alpha}
	-\frac{g}{h}\right)^{n-1} \non \\
&\times&\left[\theta\left(\bar{\alpha}-\frac{g}{h}\right)
	-\theta\left(\bar{\alpha}-\frac{g}{h}-R_\mp\alpha\right)\right]
	\frac{\solidp}{[F(\bar{\alpha},\bar{z};p,P)-i\epsilon]^{n+1}}. 
\label{eta_to_alpha_mod}
\eea

We complete our derivation of the integral equation by integrating by parts 
with respect to $\bar{\alpha}$ in order to reduce the power of $1/F(\ldots)$ 
from $n+1$ to $n$, noting that the boundary term resultant from
such an integration vanishes due to the presence of the step functions.
We therefore have, finally,

%\bea
%\Gamma^{[\ell,\ell_z]}(p,P)&=&\sum_{\rm ch}\int\limits_\Omega d\vec{\xi}\int\limits_0^\infty 
%	d\gamma \rho_{\rm ch}(\gamma,\vec{\xi})\int\limits_{0}^{\infty} 
%	d\alpha\int\limits_{-1}^{1}dz \rho_n^{[\ell]}(\alpha,z)\frac{1}{(4\pi)^2}
%	\int\limits_{-\infty}^{\infty}d\bar{\alpha}\int\limits_{-1}^{1}d\bar{z} \non \\
%&\times&\frac{\bar{\alpha}^n}{\alpha^n}{\cal K_n^{[\ell]}}(\bar{\alpha},
%	\bar{z};\alpha,z)\frac{\solidp}{[F(\bar{\alpha},\bar{z};p,P)
%	-i\epsilon]^n}
%\label{new_vertex}
%\eea

\bea
\Gamma^{[\ell,\ell_z]}(p,P)&=&\solidp\int\limits_{-\infty}^{\infty}d\bar{\alpha}
	\int\limits_{-1}^{1}d\bar{z} \frac{1}{[F(\bar{\alpha},
	\bar{z};p,P)-i\epsilon]^n}\non \\
&\times&\int\limits_{0}^{\infty} d\alpha\int\limits_{-1}^{1}dz 
	\frac{\rho_n^{[\ell]}(\alpha,z)}{\alpha^n}\;
	\bar{\alpha}^n\left[\lambda\;
	{}^{\rm tot}\!{\cal K}_n^{[\ell]}(\bar{\alpha},\bar{z};\alpha,z)\right].
\label{new_vertex}
\eea

We may use the uniqueness theorem of PTIR~\cite{Nakanishi_graph} to obtain 
the equation which we will solve numerically:

\be
\frac{1}{\lambda}
\frac{\rho_n^{[\ell]}(\bar{\alpha},\bar{z})}{\bar{\alpha}^n}=
	\int\limits_{0}^{\infty} d\alpha\int\limits_{-1}^{1}dz \;
	{}^{\rm tot}\!{\cal K}_n^{[\ell]}
	(\bar{\alpha},\bar{z};\alpha,z)
	\frac{\rho_n^{[\ell]}(\alpha,z)}{\alpha^n}\;,
\label{inteqn}
\ee
where the full analytical expression for the 
kernel function ${}^{\rm tot}{\cal K}$ can be written as
\be
{}^{\rm tot}{\cal K}={\cal K}_n^{[\ell]}(\bar{\alpha},\bar{z};0,0)-
{\cal K}_n^{[\ell]}(\bar{\alpha},\bar{z};\alpha,z),
\label{tot_kernel}
\ee
where ${\cal K}$ is
the function:
\bea
	{\cal K}_n^{[\ell]}(\bar{\alpha},\bar{z};\alpha,z)&\equiv&
	\sum_{\rm ch}\int\limits_\Omega d\vec{\xi}\int\limits_0^\infty
	d\gamma\;
	\left[\frac{1}{g^2}
	\rho_{\rm ch}(\gamma,\vec{\xi})\right]
	K_n^{[\ell]}(\bar{\alpha},
	\bar{z};\alpha,z;\gamma,\vec{\xi})\non \\
K_n^{[\ell]}(\bar{\alpha},\bar{z};\alpha,z;\gamma,\vec{\xi})
	&\equiv&\frac{1}{\bar{\alpha}^n}
	\frac{1}{|b|}\left(-\frac{b}{2}\right)^\ell\frac{\partial}
	{\partial\bar{\alpha}}\int\limits_a^\infty dy'\frac{1}{(y')^{\ell+1}}
	\left(\bar{\alpha}-\frac{g'(\bar{z},y')}{h'(y')}\right)^{n-1} \non \\
&\times&\sum_\mp \theta(\pm(G(\bar{z};y')-z))\theta(1\mp G'(\bar{z};y')) \non \\
&\times&\theta\left(\bar{\alpha}-\frac{g'(\bar{z},y)}{h'(y')}-
	\frac{y'(y'-a)}{h'(y')}\frac{1\mp G'(\bar{z};y')}{1\mp z}\alpha\right).
\label{kernel}
\eea

Note that we have made a shift of variable from $y$ to $y'$. The quantities
$g',h'$ and $G'$ are the same as their unprimed counterparts, except that
the $y$ dependence of these functions has been transformed according to
$y\rightarrow y'=y+a$. Note also that we indicate explicitly the dependence
of $K_n^{[\ell]}$ on the scattering kernel parameters $\{\gamma,\vec{\xi}\}$.
For the remainder we will omit these additional labels for brevity.

In order to implement this kernel numerically, we must perform the derivative
with respect to $\bar{\alpha}$ and simplify the resultant expression,
as well as transforming those integration variables with semi-infinite or 
infinite ranges to variables which have a finite range. We begin by
performing the $\bar{\alpha}$ derivative, which splits the kernel into
two pieces, one of which contains a delta function. After this differentiation,
we have

\bea
&&K_n^{[\ell]}(\bar{\alpha},\bar{z};\alpha,z)=
	\frac{1}{\bar{\alpha}^n}
	\frac{1}{|b|}\left(-\frac{b}{2}\right)^\ell\int\limits_a^\infty dy'
	\frac{1}{(y')^{\ell+1}} \non \\
 	&\times&\left\{(n-1)\left(\bar{\alpha}-\frac{g'}{h'}\right)^{n-2}
		\sum_\mp \theta(\pm(G'-z))\theta(1\mp G')
		\theta\left(\bar{\alpha}-\frac{g'}{h'}-\frac{y'(y'-a)}{h'}
		\frac{1\mp G'}{1\mp z}\alpha\right)\right. \non \\
	&+&\left.\left(\bar{\alpha}-\frac{g'}{h'}\right)^{n-1}
		\sum_\mp \theta(\pm(G'-z))\theta(1\mp G')
		\delta\left(\bar{\alpha}-\frac{g'}{h'}-\frac{y'(y'-a)}{h'}
		\frac{1\mp G'}{1\mp z}\alpha\right)\right\}.
\label{after-deriv}		
\eea

The piece containing the delta function may be integrated over $y'$ in a 
relatively straightforward manner, simply by rewriting the delta function 
in terms of $y'$. The argument of the delta function is quadratic in
$y'$:
\be
{\rm argument}=\frac{-1}{cy'-\frac{b^2}{4}}(A_\mp(\bar{z};\alpha,z) y'^2 
	+ B_\mp(\bar{\alpha},\bar{z};\alpha,z) y' 
	+ C(\bar{\alpha},\bar{z})).
\label{arg-delta}
\ee
The delta function is therefore
\be
\delta\left(-\frac{1}{cy'-\frac{b^2}{4}}(A_\mp y'^2 + B_\mp y' + C)\right)
	=\sum_{i=1}^2 \frac{h'(y'_i)}
	{\sqrt{D_\mp(\bar{\alpha},\bar{z};\alpha,z)}}
	\delta(y'-y'_i)\theta(D_\mp(\bar{\alpha},\bar{z};\alpha,z)),
\label{transformed_delta}
\ee
where $D_\mp\equiv B_\mp^2 - 4 A_\mp C$, and the $y'_i$ are the roots
of the quadratic, {\it i.e.}
\be
y'_1=\frac{ -B_\mp-\sqrt{D_\mp} }{2 A_\mp}, \,
y'_2=\frac{ -B_\mp+\sqrt{D_\mp} }{2 A_\mp}.
\ee

The case $n=2$ is of particular interest to us, and so we will restrict
ourselves to this case from now on. Dropping the prime on $y$,
the $n=2$ kernel function may be written as
\bea
&&K_{n=2}^{[\ell]}(\bar{\alpha},\bar{z};\alpha,z)= 
	\frac{1}{\bar{\alpha}^2}
	\frac{1}{|b|}\left(-\frac{b}{2}\right)^\ell \non \\
&\times& \sum_\mp \left\{\int\limits_a^\infty dy\: \frac{1}{y^{\ell+1}}
	\theta\left(-\frac{1}{cy-\frac{b^2}{4}}
	\left(A_\mp(\bar{z};\alpha,z)y^2+B_\mp(\bar{\alpha},\bar{z};\alpha,z)y
	+C(\bar{\alpha},\bar{z})\right)\right) \right. \non \\
&+& \left. \frac{\alpha\:\theta\left({D_\mp(\bar{\alpha},\bar{z};\alpha,z)}\right)}
	{\sqrt{D_\mp(\bar{\alpha},\bar{z};\alpha,z)}}
	\sum_{i=1}^2 \left(\frac{1\mp\frac{-2(c\bar{z}-f)}{b}}{1\mp z}
	\frac{1}{y_i^{\ell+1}}-\frac{a\mp \left(-\frac{b}{2}\bar{z}+e\right)}
	{1\mp z}\frac{1}{y_i^\ell}\right)\int\limits_a^\infty dy\:\delta(y-y_i)\right\}
	\non \\
&\times&\theta\left(\frac{1}{y-a}\left[\pm \left(\frac{-2(c\bar{z}-f)}{b}-z
	\right)y \pm \left(az-\left(-\frac{b}{2}\bar{z}+e\right)\right)\right]
	\right)	\non \\
&\times&\theta\left(\frac{1}{y-a}\left[\left(1\mp \frac{-2(c\bar{z}-f)}{b}
	\right)y -a\pm \left(-\frac{b}{2}\bar{z}+e\right)\right]\right),
\eea
where
\bea
D_\mp(\bar{\alpha},\bar{z};\alpha,z)&=&B_\mp^2(\bar{\alpha},\bar{z};\alpha,z)
	-4 A_\mp(\bar{z};\alpha,z) C(\bar{\alpha},\bar{z}) \non \\
A_\mp(\bar{z};\alpha,z)&=&m^2-\left(1-\left(\frac{-2(c\bar{z}-f)}{b}\right)^2
	\right)\qtrP + \frac{1\mp \frac{-2(c\bar{z}-f)}{b}}{1 \mp z}\alpha 
	\non \\
B_\mp(\bar{\alpha},\bar{z};\alpha,z)&=&\gamma - c\bar{\alpha} - (a+c)m^2
	+(a+c-4d-2\bar{z}(c\bar{z}-f))\qtrP - \frac{a\mp \left(-\frac{b}{2}
	\bar{z}+e\right)}{1 \mp z}\alpha \non \\
C(\bar{\alpha},\bar{z})&=&\frac{b^2}{4}(\bar{\alpha}+m^2-(1-\bar{z}^2)\qtrP).
\eea

For the purposes of numerical solution we now make successive transformations
of the integration variable $y$, first to $\tilde{y}=1/y$, and then from
$\tilde{y}$ to $Y=(b^2/4)\tilde{y}$. The first transformation serves to render
the range of integration finite, while the second ensures that we do not
encounter any difficulties in the kernel function in the limit $b\rightarrow 0$,
which can occur for example in the separable kernel case. The kernel function
after these transformations becomes
\bea
&&K_{n=2}^{[\ell]}(\bar{\alpha},\bar{z};\alpha,z) \non \\
&=&	\frac{1}{\bar{\alpha}^2}\left(-\frac{2}{b}\right)^\ell\frac{1}{|b|}
	\sum_\mp\left\{\int\limits_0^\frac{b^2}{4a}dY Y^{\ell-1}
	\theta\left(-(\widetilde{C}_\mp Y^2 + \widetilde{B}_\mp Y
	+\widetilde{A}_\mp)\right)\right. \non \\
&+& \left.\frac{\alpha\theta\left({\widetilde{D}_\mp(\bar{\alpha},\bar{z};
	\alpha,z)}\right)}
	{\sqrt{\widetilde{D}_\mp(\bar{\alpha},\bar{z};\alpha,z)}}
	\sum_{i=1}^2\left[\left(\frac{b^2}{4}\mp \tilde{g}_0\right)\frac{1}{Y_i}
	-(a\mp h_0)\right]Y_i^\ell\int\limits_0^\frac{b^2}{4a}dY \delta(Y-Y_i)\right\}
	\non \\
&\times&\theta\left(\pm (az-h_0)Y \pm (\tilde{g}_0-\frac{b^2}{4}z)\right)
	\theta\left((-a\pm h_0)Y+\frac{b^2}{4}\mp\tilde{g}_0\right),
\label{finalkernel}
\eea
where $h_0(\bar{z})=(-b/2)\bar{z}+e$, and $\tilde{g}_0(\bar{z})=(-b/2)(c\bar{z}
	-f)$. The $Y_i$ are the roots of the quadratic $\widetilde{C}_\mp Y^2
	+\widetilde{B}_\mp Y+\widetilde{A}_\mp$, and 
\bea
\widetilde{C}_\mp(\bar{\alpha},\bar{z};z)&=&(1\mp z)
	\left(\bar{\alpha}+m^2-(1-\bar{z}^2)\qtrP\right) \non \\
\widetilde{B}_\mp(\bar{\alpha},\bar{z};\alpha,z)&=&(1\mp z)
	(\gamma-c\bar{\alpha}-(a+c)m^2+(a+c-4d-2\bar{z}(c\bar{z}-f))\qtrP)
	\non \\
	&-&(a\mp (-\frac{b}{2}\bar{z}+e))\alpha \non \\
\widetilde{A}_\mp(\bar{z};\alpha,z)&=&(1\mp z)\left(\frac{b^2}{4}(m^2-\qtrP)
	+(c\bar{z}-f)^2\qtrP\right)+\left(\frac{b^2}{4}\mp \frac{-b}{2}
	(c\bar{z}-f)\right)\alpha \non \\
\widetilde{D}_\mp&=&\widetilde{B}_\mp^2 - 4\widetilde{A}_\mp\widetilde{C}_\mp
\label{tilde_fns}
\eea
This is the expression which we implement numerically. Note that the support
of the kernel is entirely determined by the step functions in 
Eq. (\ref{finalkernel}). In general it is not possible to extract the
support analytically, and so in most cases this step must be done numerically.

\section{Kernel Singularities}
\label{Singularity}

In this section we discuss the structure of the kernel function 
$K^{[\ell]}_n(\bar\alpha,\bar z;\alpha,z)$ for arbitrary $\ell$ with a fixed
kernel parameter set $(\gamma, \vec\xi)$, i.e., for constant
$\{\gamma,a_{\rm ch},b_{\rm ch},c_{\rm ch},\dots,f_{\rm ch}\}$.
We will in this section omit for brevity the subscript {\rm ch}.  
Since the case $n=2$ is of particular interest to us for numerical 
treatment, we discuss possible singularities of the kernel function 
$K^{[\ell]}_{n=2}(\bar\alpha,\bar z;\alpha,z)$, whose expression and 
derivation are given in Appendix \ref{apdx_b}.  General $n$ cases 
can be also considered in a similar manner.  

As shown in Appendix \ref{apdx_b} the kernel function 
$K^{[\ell]}_{n=2}(\bar\alpha,\bar z;\alpha,z)$ given by the 
Feynman parameter integral consists of two terms, 
one containing only step functions and another containing a delta function.  
It is convenient to make the Feynman parameter $y$ finite to discuss the 
singularities of the kernel, and so we will discuss the structure of 
the kernel function 
based on the expression (\ref{finalkernel}).  

The step function term is given by the integral
\begin{equation}
	\int_{Y_{\rm min}}^{Y_{\rm max}}dY\,Y^{\ell-1},
	\label{thetaterm}
\end{equation}
where the upper $Y_{\rm max}$ and lower limits $Y_{\rm min}$ of the integral 
are determined by relatively complicated step functions depending on 
the variables $\bar\alpha$, $\bar z$, $\alpha$, and $z$ as well as the 
kernel parameters \{$\gamma$, $a$,$\ldots$,$f$\}.  
It is easy to show that the upper limit $Y_{\rm max}$ is finite 
as long as the parameter $a$ does not vanish.  Since the scattering kernel with 
identically vanishing $a$ is nothing but the constant scattering kernel 
in the relative momentum $p$, we do not consider this case.  
Thus the Feynman parameter integral may diverge logarithmically, 
and this only 
if $Y_{\rm min}$ vanishes for the $\ell=0$ case.  However, as is clear 
from the expression (\ref{finalkernel}), the point $Y=0$ is always excluded 
by the step functions, so this integral never diverges.  

The delta function term can be written as a sum of fractions with 
square root factors in their denominator together with finite numerators.  
The square root factor comes from the 
Jacobian to change the variable of the delta function 
from the spectral variable $\bar\alpha$ to the Feynman parameter $Y$.  
%(\ref{transformed_delta})
Note that this situation is quite general and occurs for any 
angular momentum $\ell$ and dummy parameter $n$.  
From the argument of the square root, this term becomes singular
if $\alpha$ satisfies the following quadratic equation:
\begin{eqnarray}
 & &\left( \frac{a\mp h_{0}(\bar z)}{1\mp z} \alpha - B(\bar\alpha,\bar z) 
 -  2 \frac{b^{2}/4\mp \tilde g_{0}(\bar z)}{a\mp h_{0}(\bar z)}
 C(\bar\alpha,\bar z) \right)^{2}  \nonumber \\
 &  & \qquad -4 C(\bar\alpha,\bar z) \left( A(\bar z) + B(\bar\alpha,\bar z) 
 \frac{b^{2}/4\mp \tilde g_{0}(\bar z)}{a\mp h_{0}(\bar z)} 
 + C(\bar\alpha,\bar z) 
 \left( \frac{b^{2}/4\mp \tilde g_{0}(\bar z)}{a\mp h_{0}(\bar z)} 
 \right)^{2} \right) =0,
\label{argofsqrt}
\end{eqnarray}
where $h_0(\bar{z})=(-b/2)\bar{z}+e$, 
and $\tilde{g}_0(\bar{z})=(-b/2)(c\bar{z}-f)$.  The functions 
$A(\bar z)$, $B(\bar\alpha,\bar z)$ and $C(\bar\alpha,\bar z)$ are
\begin{eqnarray}
  A(\bar z) & = & \frac{b^{2}}{4}\left(m^2-\frac{P^{2}}{4}\right)+
  (c\bar{z}-f)^2\frac{P^{2}}{4},
	\nonumber  \\
  B(\bar\alpha,\bar z) & = & \gamma-c\bar{\alpha}-(a+c)m^2+
  (a+c-4d-2\bar{z}(c\bar{z}-f))\frac{P^{2}}{4},
	\nonumber \\
	C(\bar\alpha,\bar z) & = & \bar{\alpha}+m^2-(1-\bar{z}^2)\frac{P^{2}}{4}.
	\label{abc}
\end{eqnarray}
Thus the kernel function diverges as a square root if Eq. (\ref{argofsqrt}) 
possesses a simple root.  On the other hand, the kernel function diverges 
linearly if Eq. (\ref{argofsqrt}) admits a double root.  
Since $C(\bar\alpha,\bar z) > 0$ for any bound state, 
a double root occurs only if the terms in the second set of parentheses 
cancel.  In this case, however, the residue of this pole (linear 
singularity) vanishes, so that the delta function term stays finite as a whole.

To summarise: the kernel function 
$K^{[\ell]}_{n=2}(\bar\alpha,\bar z;\alpha,z)$ for
a fixed kernel parameter set $\{\gamma,a_{\rm ch},b_{\rm ch},\dots,
f_{\rm ch}\}$ contains only integrable 
square root singularities at the boundary of its support, which if appropriately
treated numerically present no difficulties.

%=======================================================================
%          Bibliography:
%-----------------------------------------------------------------------
%%%%%%%%%%%%%%%%%

%%%%%%%%%%%%%%%%%
%=======================================================================
%        tables:
%-----------------------------------------------------------------------
%
%
\begin{table}[hbt]
\begin{center}
\begin{tabular}{lrr|lrr} \hline
$\eta$ & $\lambda_{E}$ & $\lambda$ & $\eta$ & $\lambda_{E}$ & $\lambda$ 
\\ \hline
0 & 2.5658    & 2.5662 & 0.8 & 1.4055    & 1.4056 \\ 
0.2 & 2.4984    & 2.4988 & 0.9 & 1.0349    & 1.0350 \\ 
0.4 & 2.2933    & 2.2937 & 0.99 & 0.5167    & 0.5168 \\ 
0.6 & 1.9398    & 1.9402 & 0.999 & 0.3852    & 0.3853 \\ 
\end{tabular}
\parbox{130mm}{\caption{
Comparison of the coupling strengths, $\lambda\equiv g^2/(4\pi)^2$, obtained
for the ladder approximation kernel from the Euclidean ({\it i.e.}, 
Wick-rotated) $s$-wave solution ($\lambda_E$) and those obtained here
directly in Minkowski space ($\lambda$), using a moderate grid choice for
$\alpha$ and $z$. The Wick-rotated values are from
Linden and Mitter~\protect{\cite{L+M}}. 
The parameter $\eta$ is the ``fraction of
binding'', $\eta\equiv\protect{\sqrt{P^2}}/2m=M/2m$, where $P^2=M^2$
and $M$ is the mass of the bound state ({\it i.e.}, $0\leq\eta<1$).
The results shown here had $m_\sigma=m/2$ for the exchange particle
($\sigma$) mass.
}
\label{table_pure_ladder}}
\end{center}
\end{table}
%=======================================================================
%        figures:
%-----------------------------------------------------------------------

\begin{thebibliography}{99}
%%%%%%%%%%%%%%%%%
%
\bibitem{BS_amplitude} K.~Kusaka and A.G.~Williams, {\it Phys. Rev.} {\bf D51},
 7026 (1995).
%
\bibitem{Nakanishi_survey} N.~Nakanishi, {\it Suppl. Prog. Theor. Phys.}
{\bf 43}, 1 (1969).
%``Review of the Wick-Cutkosky Model''
%\bibitem{Nakanishi_WC} N.~Nakanishi, {\it Suppl. Prog. Theor. Phys.} {\bf 95},
%1 (1988).
%
%``Review of the Spinor-Spinor Bethe-Salpeter Equation: Spectral Properties
%  in the Massless Ladder Model''
%\bibitem{Seto} N.~Seto, {\it Suppl. Prog. Theor. Phys.} {\bf 95}, 25 (1988).
%
%%``Hyperfine Structure in Positronium''
%\bibitem{Murota} T.~Murota, {\it Suppl. Prog. Theor. Phys.} {\bf 95}, 46
%(1988).
%
%``Bibliography of the Bethe-Salpeter Equation''
%\bibitem{BSE_refs} T.~Murota {\it et al.}, {\it Suppl. Prog. Theor. Phys.}
{\bf 95}, 78 (1988).
%
\bibitem{B+D} J.D.~Bjorken and S.D.~Drell, {\it Relativistic Quantum Fields}.
McGraw-Hill, New York, 1965.
%
\bibitem{I+Z} C.~Itzykson and J.-B.~Zuber, {\it Quantum Field Theory}.
	McGraw-Hill, New York, 1980.
%
\bibitem{TheReview}  C.~D.~Roberts and A.~G.~Williams,
    {\it Dyson-Schwinger Equations and their Application to Hadronic
    Physics\/}, in
    {\it Progress in Particle and Nuclear Physics, Vol.~33}
    (Pergamon Press, Oxford, 1994), p.~477.
%
%\bibitem{Wick} G.C.~Wick, {\it Phys. Rev.} {\bf 96}, 1124 (1954).
%
\bibitem{Nakanishi_graph} N.~Nakanishi, {\it Graph Theory and Feynman
Integrals}. Gordon and Breach, New York, 1971.
%
%\bibitem{Cutkosky} R.E.~Cutkosky, {\it Phys. Rev.} {\bf 96}, 1135 (1954).
%
%\bibitem{Wanders} G.~Wanders, {\it Helv. Phys. Acta} {\bf 30}, 417 (1957).
%
%\bibitem{I+M} M.~Ida and K.~Maki, {\it Prog. Theor. Phys.} {\bf 26},
%470 (1961).
%
\bibitem{Nakanishi63} N.~Nakanishi, {\it Phys. Rev.} {\bf 130}, 1230 (1963);
Erratum, {\it ibid} {\bf 131}, 2841 (1963).
%
\bibitem{Sato} I.~Sato, {\it J. Math. Phys.} {\bf 4}, 24 (1963).
%
%\bibitem{Fainberg} V.~Ya.~Fa\v \i nberg, {\it J.E.T.P.} {\bf 36}, 1503 (1959)
%[{\it Sov. Phys. J.E.T.P.}{\bf 9}, 1066 (1959)].
%%
%\bibitem{D+G+S} S.~Deser, W.~Gilbert, and E.C.G.~Sudarshan, {\it Phys. Rev.}
%{\bf 115}, 731 (1959).
%
%\bibitem{Ida} M.~Ida, {\it Prog. Theor. Phys.} {\bf 23}, 1151 (1960).
%
\bibitem{L+M} E.~zur Linden and H.~Mitter, {\it Nuovo Cim.} {\bf 61B}, 
389 (1969).
%
\bibitem{Tjon1} T.~Nieuwenhuis and J.A.~Tjon, {\it Phys. Rev. Lett.} {\bf 77}, 
814 (1996).
%
\bibitem{Tjon2} T.~Nieuwenhuis and J.A.~Tjon, {\it Few-Body Syst.} {\bf 21},
167 (1996).
%
%%%%%%%%%%%%%%%%%
\end{thebibliography}
\end{document}